\documentclass[12pt,letterpaper,english]{article}

 \usepackage{setspace,graphics}
 \usepackage[dvips]{epsfig} 
 \usepackage{a4,amssymb,epsfig,array,cite}
 \usepackage[hypertex]{hyperref}

\usepackage{amsfonts,bm,amssymb,euscript,array,babel}

\newcounter{multieqs}

\newcommand{\be}{\begin{equation}}
\newcommand{\ee}{\end{equation}}
\newcommand{\eq}[1]{(\ref{#1})}

\def\nn{\nonumber}
\def\bea{\begin{eqnarray}}
\def\eea{\end{eqnarray}}

\def\beqa{\begin{eqnarray}} 
\def\eeqa{\end{eqnarray}} 
\def\beq{\begin{equation}} 
\def\eeq{\end{equation}}

\def\a{\alpha}          
\def\b{\beta}           
    
\def\d{\delta}

\def\g{\gamma}

\def\l{\lambda} \def\L{\Lambda} \def\la{\lambda}

\def\s{\sigma}

\def\cA{{\cal A}}  \def\cC{{\cal C}}
  \def\cF{{\cal F}}
\def\cG{{\cal G}} \def\cH{{\cal H}} 
\def\cJ{{\cal J}}  
\def\cM{{\cal M}} \def\cN{{\cal N}} 
  
  \def\cU{{\cal U}}

\def\R{{\mathbb R}}

\def\one{\mbox{1 \kern-.59em {\rm l}}}

\def\bit{\begin{itemize}}
\def\eit{\end{itemize}}

\def\({\left(}
\def\){\right)}
\def\diag{\mbox{diag}}

\def\d{\delta}

\def\uno{\mbox{1 \kern-.59em {\rm l}}}

\newcommand{\tr}{\mbox{tr}}

\def\bcomment#1{}

\renewcommand{\title}[1]{\vspace{10mm}\noindent{\Large{\bf #1}}\vspace{8mm}}
\newcommand{\authors}[1]{\noindent{\large #1}\vspace{5mm}}
\newcommand{\address}[1]{{\itshape #1\vspace{2mm}}}

\begin{document}


\begin{flushright}
UWTHPh-2008-20\\
\end{flushright}

\begin{center}

\title{Covariant Field Equations, Gauge Fields and  \\[1ex] Conservation Laws
 from Yang-Mills Matrix Models}

\authors{Harold {\sc Steinacker}${}^{1}$}

\address{ Fakult\"at f\"ur Physik, Universit\"at Wien\\
Boltzmanngasse 5, A-1090 Wien, Austria}

\footnotetext[1]{harold.steinacker@univie.ac.at}

\vskip 1.5cm

\textbf{Abstract}

\vskip 3mm 

\begin{minipage}{14cm}%

The effective geometry and the gravitational coupling 
of nonabelian gauge and scalar fields on generic NC branes in
Yang-Mills matrix models is determined.
Covariant field equations are derived from the basic 
matrix equations of motions, known as Yang-Mills algebra. 
Remarkably, the equations of motion for the Poisson structure 
and for the nonabelian gauge fields
follow from a matrix Noether theorem,
and are therefore protected from quantum corrections.
This provides a transparent derivation and generalization of the 
effective action governing the $SU(n)$ gauge fields  obtained in [1], 
including the would-be topological term.
In particular, the IKKT matrix model is capable of
describing 4-dimensional NC space-times with 
a general effective metric.
Metric deformations of flat Moyal-Weyl space are briefly discussed.

\end{minipage}

\end{center}


\setcounter{page}0
\thispagestyle{empty}
\newpage

\begin{spacing}{.3}
{
\noindent\rule\textwidth{.1pt}            
   \tableofcontents
\vspace{.6cm}
\noindent\rule\textwidth{.1pt}
}
\end{spacing}


\section{Introduction}

This paper is a continuation of a series of papers 
\cite{Steinacker:2007dq,Grosse:2008xr,Klammer:2008df,Steinacker:2008ri}
with the aim to understand the geometry and the physics 
on noncommutative (NC) branes in Matrix-Models
of Yang-Mills type. The motivation of this enterprise is to 
study whether such branes can serve as
dynamical space-time including matter, gauge fields and gravity. 
In the previous papers, it was shown that these matrix models
provide indeed a mechanism for emergent gravity on the branes, with
an effective metric which couples universally (up to possibly
conformal factors) to bosonic and fermionic matter as well as
nonabelian gauge fields. The mechanism of gravity 
on these branes is 
different from general relativity a priori, and related to
gauge theory on non-commutative spaces. 
This is very similar in spirit 
to the ideas put forward in 
\cite{Rivelles:2002ez,Yang:2006mn,Yang:2008fb}; 
see also 
e.g. \cite{Muthukumar:2004wj,Banerjee:2004rs,Fatollahi:2008dg} for related work.

The starting point of our approach are
matrix models of Yang-Mills type, with bosonic action
\be
S_{YM} = - Tr [X^a,X^b] [X^{a'},X^{b'}] \eta_{aa'}\eta_{bb'}
\label{YM-action-0}
\ee
where $\eta_{ab}$ is a constant background metric with Euclidean or
Minkowski signature. These models arise e.g. through 
dimensional reduction of $U(\cN)$ Yang-Mills gauge theory 
with $\cN \to \infty$ to zero dimensions.
The models describe in particular 
noncommutative (NC) branes of even dimension $2n$, 
including e.g. the 
Moyal-Weyl quantum plane $[X^\mu,X^\nu] = i \theta^{\mu\nu} \one$
which is a solution of the equations of motion. 
It was generally believed that fluctuations of the matrices or
``covariant coordinates'' $X^a$ describe
gauge fields on such NC spaces.
However, a detailed analysis 
\cite{Steinacker:2007dq,Rivelles:2002ez} shows that 
the $U(1)$ components of these fluctuations are actually 
gravitational degrees of freedom, and only the $SU(n)$ 
components should be interpreted as nonabelian gauge fields. 
This provides a detailed explanation of 
UV/IR mixing in NC gauge theory in terms of an induced gravitational
action \cite{Grosse:2008xr}. 
The effective action 
for the $SU(n)$ gauge fields was found in 
\cite{Steinacker:2007dq}  using a 
 brute-force computation. It was shown that 
 the gauge fields couple as expected to the effective gravitational 
 metric, and an interesting
``would-be topological'' term in the action was 
found.

The effective metric on the NC branes
was generalized in \cite{Steinacker:2008ri} to the case of 
non-trivially embedded branes $\cM_\theta \subset \R^D$ in 
higher-dimensional matrix models, and 
has the form $\tilde G^{\mu\nu} = e^{-\sigma} \theta^{\mu\mu'}(x) \theta^{\nu\nu'}(x) 
 g_{\mu'\nu'}(x)$. This is very similar to the open string metric
in the context of string theory \cite{Seiberg:1999vs}, while
the ``embedding metric'' $g_{\mu'\nu'}(x)$ corresponds to the
closed string metric and enters the action on the branes 
only indirectly.
Indeed it appears that only $D>4$ can provide 
a large enough class of effective metrics to 
allow a realistic description of gravity.
The case $D=10$ is favored from 
the point of view of quantization, more precisely the 
IKKT model \cite{Ishibashi:1996xs} 
which was introduced originally in the context
of string theory; this  allows to describe the most general
effective metric on 4-dimensional branes.
However, the coupling of nonabelian gauge fields 
to this general metric was not established up to now.

In the present paper, we study in detail gauge fields on generic
NC branes $\cM_\theta \subset \R^D$ with $D>4$  
governed by the matrix model \eq{YM-action-0}.
$SU(n)$ gauge fields are shown to arise on $\cM_\theta$ 
as non-abelian fluctuations of the matrices or
``covariant coordinates'' $X^a$, and couple as 
expected to the effective metric
$\tilde G^{\mu\nu}$.
We provide a new derivation of the Yang-Mills 
equations of motion \eq{eom-nonabel}
for nonabelian gauge fields 
coupled to $\tilde G^{\mu\nu}$ on general NC branes, 
as well as the equation governing the noncommutativity resp. Poisson
tensor $\theta^{\mu\nu}(x)$.
Remarkably, both equations can be understood as 
consequences of a basic ``translational'' symmetry of the underlying matrix
model $X^a \to X^a + c^a \one$. The corresponding Noether theorem 
yields precisely the above equations. This provides not only a
transparent derivation of these equations of motion;
as usual in quantum field theory, a derivation 
based on a fundamental symmetry
also implies that these equations are protected from quantum
corrections.  Therefore these equations of motion 
are valid also at the quantum level. 
In particular, \eq{eom-geom-covar-extra} 
provides the relation between non-commutativity 
of space-time and gravity.

Form a mathematical point of view, it is quite remarkable that
the  covariant Yang-Mills equations 
\eq{eom-nonabel} as well as 
\eq{eom-geom-covar-extra} are direct consequences of the 
underlying matrix equations of motion
known as Yang-Mills algebra \cite{Nekrasov:2002kc,Connes:2002ya},
\be
[X^a,[X^b,X^{a'}]] g_{aa'} =0
\label{YM-matrix-eom}
\ee
in the background of a quantized Poisson manifold.
The resulting structure is very rigid and leaves little room for 
different interpretations. Combined with the
well-known fact \cite{friedman} that any analytic
4-dimensional (pseudo) Riemannian manifold $(\cM^4,g_{\mu\nu})$ 
can be embedded locally isometrically in
flat (pseudo)Euclidean space $\R^{10}$, it follows that 
the Riemannian geometry
required for gravity can be described in terms of a simple matrix
model of the above type, notably the IKKT model 
with $D=10$. These models therefore provide a possible
foundation of geometry and gravity, without any 
classical-geometrical prerequisites: the geometry arises only effectively
in a semi-classical limit.
Moreover, gravitons are  naturally unified with 
(nonabelian) gauge fields, both being described by fluctuations 
of the underlying matrices around some given background. 
This provides a simple and promising framework for a unified description
for gauge fields, gravity and matter.
Moreover, it is reasonable to expect that the IKKT model 
is well-defined at quantum level, due to its extended supersymmetry.
This supports the hypothesis 
that these matrix models may be a
suitable framework to describe the quantum regime of gravity.


Some new results on various aspects of the geometry
of the NC branes are also given. 
In particular, 
we elaborate in some detail the preferred frame and gauge 
defined by the matrix model. This provides a 
simplified derivation of the fact that
the linearized metric fluctuations around flat Moyal-Weyl 
space due to the would-be $U(1)$ gauge fields are 
Ricci-flat but non-trivial, as first observed by 
Rivelles \cite{Rivelles:2002ez}; they
are hence interpreted as gravitons. 
We give explicit expressions for the 
linearized Riemann tensors, including also fluctuations of the 
embedding.
 Nevertheless, the physical aspects of the emergent gravity 
theory are not sufficiently well understood 
at this point and require much more work.

The results of this paper illustrate the unexpected role of symmetries 
in emergent gravity. The classical concepts of diffeomorphism
resp. general coordinate invariance loose their significance, 
being partially replaced by symplectomorphism invariance; 
see also \cite{Yang:2008fb} for related discussion 
from a somewhat different point of view. 
On the other hand, novel  symmetries such as the above ``translational''
symmetry arise with unexpected consequences. 
In particular, we point out that the global $SO(D)$ symmetry of the 
model can be exploited by going to certain 
``normal embedding coordinates'' where some of the analysis simplifies.

There is of course a relation with previous work in the context of 
matrix models in string theory
(e.g.\cite{Banks:1996vh,Banks:1996nn,Ishibashi:1996xs,Chepelev:1997ug,Aoki:1998vn,Aoki:1999vr,Nair:1998bp,Myers:1999ps,Azuma:2004qe,Azuma:2005pm,Taylor:2001vb})
and NC gauge theory
(e.g.\cite{Alekseev:2000fd,Grosse:2004wm,Behr:2005wp}). In particular,
evidence for gravity on NC branes was obtained
e.g. in \cite{Ishibashi:2000hh,Kimura:2000ur,Kitazawa:2006pj}.
However, most of this work is focused on BPS 
or highly symmetric (typically ``fuzzy'') brane solutions.
The essential point of the present approach is to study 
generic curved NC branes and their effective
geometry. The proper separation of 
gravitational and gauge degrees of freedom on these branes 
in the matrix model was
understood only recently. Moreover, it turns out that 
all fields which arise due to fluctuations around such a background
{\em only} live on the brane $\cM^{2n}_\theta\subset \R^D$;
there really is no D-dimensional ``bulk''
which could carry physical degrees of freedom, not even gravitons.
Therefore the present framework is quite different from the
braneworld-scenarios such as \cite{ArkaniHamed:1998rs,Randall:1999vf}.
One could argue that the strength of
string theory (notably the good behavior under quantization) 
are preserved while the main problems
(lack of predictivity) are avoided here. Clearly all 
essential ingredients for physics are available, 
but it remains to be seen whether realistic physics
can be described through these models.
Finally, even though we focus on the semi-classical
limit, higher-order corrections in $\theta$ could and should be computed
eventually. This will provide a link with NC field theory 
\cite{Doplicher:1994tu,Douglas:2001ba,Szabo:2001kg} and possibly with
other approaches to NC gravity, 
see e.g. \cite{Aschieri:2005yw,Szabo:2006wx,Madore:2000aq}.

This paper is organized as follows. 
We start  in section \ref{sec:basic} by recalling
the basic aspects of NC 
branes in matrix models under considerations.
The frame which arises naturally is discussed in some detail in 
section \ref{sec:frame}. A preferred class of coordinates is 
discussed in section \ref{sec:SOD}, taking advantage of the underlying 
$SO(D)$ symmetry of the model.
Metric fluctuations of flat Moyal-Weyl space are discussed in 
section \ref{sec:linearized}
In section \ref{sec:noether}, the basic Noether theorem
based on the translational symmetry is exploited for the 
$U(1)$ sector. Nonabelian gauge fields are finally introduced
in section \ref{sec:nonabel}, and the effective action is 
established based on Noether's theorem.

\section{The Matrix Model}
\label{sec:basic}

Consider first the basic matrix model 
\be
S_{YM} = - Tr [X^\mu,X^\nu] [X^{\mu'},X^{\nu'}] 
\eta_{\mu\mu'}\eta_{\nu\nu'},
\label{YM-action-1}
\ee
for
\be
\eta_{\mu\mu'} = \diag(1,1,1,1) \quad \mbox{or}
 \quad \eta_{\mu\mu'} = \diag(-1,1,1,1) 
\label{background-metric}
\ee
in the Euclidean\footnote{The reason for this notation is
  to avoid using $g_{\mu\nu}$
for this fixed ``metric''; $g_{\mu\nu}$ will arise later.}  resp.  Minkowski case.
The  $X^\mu$ for $\mu=1,2,3,4$ are hermitian matrices ("covariant coordinates")
or operators acting on some Hilbert space $\cH$.
The above action is invariant under the following fundamental gauge symmetry
\be
X^\mu \to U^{-1} X^\mu U, \qquad U \in \cU(\cH) .
\label{gauge}
\ee
We will denote the commutator of 2 matrices as 
\be
[X^\mu,X^\nu] = i \theta^{\mu\nu}
\label{theta-def}
\ee
where $\theta^{\mu\nu} \in L(\cH)$ is 
{\em not} necessarily proportional to $\one_\cH$.
We focus here on configurations $X^\mu$ 
 which can be interpreted as quantizations 
of coordinate functions $x^\mu$ on
a Poisson manifold $(\cM,\theta^{\mu\nu}(x))$ with general
Poisson structure $\theta^{\mu\nu}(x)$. 
The simplest case is the Moyal-Weyl quantum plane and its deformations
discussed in section \ref{sec:moyal-weyl}, but
essentially any Poisson manifold provides (locally) a possible
background $X^\mu$ \cite{Kontsevich:1997vb}.
Formally, this means that there is an
isomorphism\footnote{operator-technical subtleties concerning well-behaved subalgebras
etc. are not important here and will be ignored. In particular,
$x^\mu$ resp. $X^\mu$ are considered as elements of $\cC(\cM)$
resp. $\cA$} 
of vector spaces
\be
\begin{array}{rcl}
\cC(\cM) &\to& \cA\,\subset \, L(\cH)\, \\
 f(x) &\mapsto& \hat f(X) \\
\mbox{such that}\quad i\{f,g\} &\mapsto& [\hat f,\hat g] + O(\theta^2)
\end{array}
\label{map} 
\ee
Here $\cC(\cM)$ denotes some space of functions on $\cM$, 
and $\cA$ is interpreted as quantized algebra of 
functions on $\cM$. 

In this paper we only consider the {\em semi-classical} or
{\em geometrical\/} limit of such a quantum space. This means that 
the space is described in terms of functions on $\cM$ using \eq{map},
keeping only the Poisson bracket on the rhs of \eq{map} and
dropping all higher-order terms in $\theta$. 
We will accordingly replace
$[\hat f(X),\hat g(X)] \to i  \{f(x),g(x)\}$ and 
$[X^\mu,X^\nu] \to i\theta^{\mu\nu}(x)$.
In particular, 
\be
[X^\mu,f(X)] \sim i\theta^{\mu\nu}(x) 
\frac{\partial}{\partial x^\nu} f(x)
\label{derivation}
\ee
which will be used throughout this paper, denoting with
$\sim$  the leading contribution in a semi-classical 
expansion in powers of $\theta^{\mu\nu}$.

\subsection{Noncommutative branes and extra dimensions}
\label{sec:extradim}

We now review the generalization to NC branes in
higher-dimensional matrix models as discussed in \cite{Steinacker:2008ri}.
The basic tool to understand the physics is to add a scalar (``test'')
field to the model.
Using the above assumptions and preserving the 
gauge symmetry \eq{gauge}, a scalar fields can be coupled to the above
matrix model by adding an action of the form 
\be
Tr [X^\mu,\phi] [X^\nu,\phi] \eta_{\mu\nu}\, 
\,\,\sim\,\,  - Tr  \eta_{\mu\nu}\theta^{\mu\mu'}\theta^{\nu\nu'}
 \partial_\mu \phi \partial_\nu \phi . 
\ee
The combined action has the same form as the matrix model \eq{YM-action-1}
if we consider $\phi \equiv X^5 $ as extra coordinate,
but insisting that $\phi=\phi(X^\mu)$ is a function of the other 4
coordinates. This motivates to consider more generally 
\be
S_{YM} = - Tr [X^a,X^b] [X^{a'},X^{b'}] 
\eta_{aa'}\eta_{bb'},
\label{YM-action-extra}
\ee
for hermitian matrices 
or operators  $X^a,  \,\, a=1,..., D$ 
acting on some Hilbert space $\cH$, and equations of motion
\be
[X^a,[X^b,X^{a'}]] \eta_{aa'} =0 .
\label{matrix-eom}
\ee
A scalar field can therefore be interpreted
as defining an embedding of a 4-dimensional manifold 
(a ``3-brane'')
in a higher-dimensional space.  
This  suggests to consider a higher-dimensional version of the
Yang-Mills matrix model, such as the IKKT model with $D=10$.

We now to consider generic $2n$ -dimensional 
noncommutative spaces $\cM_\theta^{2n} \subset \R^{D}$
(a $2n-1$ brane), interpreted as (Euclidean or Minkowski) space-time
embedded in $D$ dimensions. $2n\geq 4$ is allowed in order to cover configurations
such as $\cM_\theta^{2n} \cong \cM^4_\theta \times K_\theta \subset
\R^D$; however we focus mainly on the case $2n=4$.
To realize this,  consider the matrix model \eq{YM-action-extra}
and split\footnote{For
generic embeddings, the separation \eq{extradim-splitting}
is arbitrary, and we are free to choose  different $2n$ components
among the $\{X^a\}$ as generators of tangential vector fields.
This  corresponds
to a very interesting transformation exchanging
fields with coordinates, related to the discussion in section
\ref{sec:SOD}.}
 the matrices as 
\be
X^a = (X^\mu,\phi^i), \qquad \mu = 1,...,2n, 
\,\,\, i=1, ..., D-2n
\label{extradim-splitting}
\ee
where the ``scalar fields'' $\phi^i=\phi^i(X^\mu)$ are assumed to 
be functions of $X^\mu$.
The basic example is a flat embedding of a
4-dimensional NC background
\bea
[X^\mu,X^\nu] &=& i\theta^{\mu\nu}, \qquad \mu, \nu= 1,...,4, \nn\\
\phi^i &=& 0,\qquad\quad\,  i=1, ..., D-4
\eea
where $X^\mu$ generates a 4-dimensional NC space $\cM^4_\theta$. 
In the semi-classical limit, suitable ``optimally localized states''
of $\langle X^a\rangle \sim x^a$  
will then be located on $\cM^{2n} \subset \R^{D}$.
One could  interpret $\phi^i(x)$ as scalar fields on 
$\R^4_\theta$ generated by $X^\mu$; however, it is more appropriate to
interpret $\phi^i(x)$ as purely geometrical degrees of freedom,
defining 
the embedding of a $2n$ - dimensional submanifold $\cM^{2n} \subset
\R^{D}$. 
This $\cM^{2n}$ carries the induced metric 
\be
g_{\mu\nu}(x) = \eta_{\mu\nu} 
+  \partial_{\mu}\phi^i \partial_{\nu}\phi^j\delta_{ij} 
\, = \, \partial_\mu x^a\partial_\nu x^b\, \eta_{ab} 
\label{g-def}
\ee
via pull-back of $\eta_{ab}$. Note that $g_{\mu\nu}(x)$
is {\em not} the metric responsible for gravity, and will enter the
action only implicitly.  
We will see below that all fields coupling to such a background will {\em only}
live on the brane $\cM^{2n}$, and there really is no higher-dimensional ``bulk''
which could carry physical degrees of freedom, not even gravitons.
Therefore the present framework is quite different from the
braneworld-scenarios such as \cite{ArkaniHamed:1998rs,Randall:1999vf}.

\paragraph{Poisson and metric structures.}

Expressing the $\phi^i$ in terms of $X^\mu$, we obtain
\be
[\phi^i,f(X^\mu)] \,\sim\, i\theta^{\mu\nu}\partial_\mu \phi^i \partial_\nu f
= ie^{\mu}(f)\, \partial_\mu \phi^i
\ee
in the semi-classical limit.
This involves only the components $\mu = 1, ..., 2n$ of the
antisymmetric ``tensor'' $[X^a,X^b] \sim i\theta^{ab}(x)$, which has rank $2n$
in the semi-classical limit.
Here 
\be
e^{\mu} := -i[X^\mu,.] \,\sim\, \theta^{\mu\nu} \partial_\nu 
\ee
are derivations, which span the tangent space of $\cM^{2n}\subset \R^{D}$.
They define a preferred frame discussed in section \ref{sec:frame}. We can interpret 
\be
[X^\mu,X^\nu] \sim i\theta^{\mu\nu}(x)
\label{theta-induced}
\ee
as Poisson structure on $\cM^{2n}$, noting that the Jacobi identity
is trivially satisfied. This is indeed the Poisson structure 
 on $\cM^{2n}$ whose quantization is given by the matrices 
$X^\mu, \,\, \mu = 1,..., 2n$, interpreted as quantization of 
the  coordinate functions $x^\mu$ on $\cM^{2n}$. 
Conversely, any $\theta^{\mu\nu}(x)$ 
can be (locally) quantized as \eq{theta-induced}, 
and provides together with arbitrary ``embedding'' functions
$\phi^i(x)$ a quantization of $\cM^{2n} \subset \R^D$ 
as described above.
In particular, the rank of $\theta^{\mu\nu}$ coincides with the 
dimension of $\cM^{2n}$.
We denote its inverse matrix with
\be
\theta^{-1}_{\mu\nu}(x)\, ,
\ee
which defines a symplectic form on $\cM^{2n}$. Then
the trace is given semi-classically by the 
volume of the symplectic form,
\bea
(2\pi)^{n}\, Tr f &\sim& \int d^{2n} x\, \rho(x)\, f \nn\\
\rho(x) &=& (\det\theta^{-1}_{\mu\nu})^{1/2} .
\label{rho-def-general}
\eea
We can now  extract the semi-classical limit
of the matrix model and its physical interpretation.
To understand the effective geometry on $\cM^{2n}$,
consider again a (test-) particle on $\cM^{2n}$, 
modeled by some additional
scalar field $\varphi$
(this could be e.g. $su(k)$ components of $\phi^i$).
The kinetic term must have the form
\bea
S[\varphi] &\equiv& - Tr [X^a,\varphi][X^b,\varphi] \eta_{ab} = 
- Tr \([X^\mu,\varphi][X^\nu,\varphi] \eta_{\mu\nu} 
  + [\phi^i,\varphi][\phi^j,\varphi] \delta_{ij}\) 
\label{MM-action-scalar}
\eea
which in the semi-classical limit can be written as  
\bea
S[\varphi]
&\sim&  \frac 1{(2\pi)^n}\,\int d^{2n} x\; \rho(x) \, G^{\mu\nu}(x)
 \partial_{\mu} \varphi \partial_{\nu} \varphi  
\label{covariant-action-scalar-0} \nn\\
 &=&\frac 1{(2\pi)^n}\, \int d^{2n} x\; 
|\tilde G_{\mu\nu}|^{1/2}\,\tilde G^{\mu\nu}(x)
 \partial_{\mu} \varphi \partial_{\nu} \varphi  \,.
\label{covariant-action-scalar}
\eea
which has the correct covariant form, where \cite{Steinacker:2008ri}
\be  \fbox{$\quad
\begin{array}{rrl}
G^{\mu\nu}(x) &=& \theta^{\mu\mu'}(x) \theta^{\nu\nu'}(x) 
 g_{\mu'\nu'}(x)   \\[1ex]
\tilde G^{\mu\nu}(x) &=& e^{-\sigma}\, G^{\mu\nu}(x) 
\label{G-tilde-def}, \\[1ex]
\rho\, G^{\mu\nu} &=& |\tilde G_{\mu\nu}|^{1/2}\,\tilde G^{\mu\nu}(x) .
\end{array} \quad  $}
\label{G-def-general}
\ee
Here $g_{\mu\nu}(x)$ 
is the metric \eq{g-def} induced on $\cM^{2n}\subset \R^{D}$ via 
pull-back 
of $\eta_{ab}$.
This amounts to
\bea
\qquad \rho &=& |\tilde G_{\mu\nu}|^{1/2}\,e^{-\sigma} , 
\label{rho-g-sigma}\\
e^{-(n-1)\sigma} &=& |G_{\mu\nu}|^{1/4}|g_{\mu\nu}(x)|^{-\frac 14}\, 
   = \rho\, |g_{\mu\nu}(x)|^{-\frac 12}  \label{sigma-rho-relation}
\eea
where we exclude the case $n=1$ for simplicity.
Therefore the kinetic term on  $\cM^{2n}_\theta$
is governed by the metric $\tilde G_{\mu\nu}(x)$, 
which depends on the Poisson tensor $\theta^{\mu\nu}$ and the 
embedding (``closed string'') metric $g_{\mu\nu}(x)$.
Similarly,
the matrix model action \eq{YM-action-extra} 
can be written in the semi-classical limit as 
\be
S_{YM} = - Tr[X^a,X^b][X^{a'},X^{b'}] \eta_{aa'} \eta_{bb'} 
\,\sim\, \frac 4{(2\pi)^n}\,\int d^{2n} x\,  \rho(x) \eta(x) ,
\label{S-semiclassical-general}
\ee
where 
\bea
\eta(y) &=& \frac 14 G^{\mu\nu}(x) g_{\mu\nu}(x) 
\sim - \frac 14 [X^a,X^b][X^{a'},X^{b'}] \eta_{aa'} \eta_{bb'} .
\label{eta-def}
\eea
We note that 
\be
|\tilde G_{\mu\nu}(x)| = |g_{\mu\nu}(x)|, \qquad 2n=4
\label{G-g-4D}
\ee
on 4-dimensional branes, hence $\tilde G_{\mu\nu}$ is
unimodular for $D=4$ in the preferred matrix coordinates
\cite{Steinacker:2007dq}. This
means that the Poisson tensor $\theta^{\mu\nu}$ does 
not enter the Riemannian volume at all. This leads to a
very interesting mechanism for ``stabilizing flat space'', 
which may hold the key for the cosmological constant problem
as discussed in \cite{Steinacker:2008ri}.

\paragraph{Some properties.}

The metric  $\tilde G^{\mu\nu}$ satisfies the following 
useful identities:
\bea
0 &=& 
\partial_\mu (\rho\,\theta^{\mu\nu}) 
= \partial_{\mu}(e^{-\sigma}\sqrt{|\tilde G|} \theta^{\mu\nu'}) 
= \sqrt{|\tilde G|}\,\tilde\nabla_{\mu} 
(e^{-\sigma} \theta^{\mu\nu'}) 
\label{partial-theta-id}\\
\tilde\Gamma^\mu &=& - \frac 1\rho e^{-\sigma}\,
\partial_\nu (G^{\nu\mu}\,\rho)
= - e^{- \sigma}\,\theta^{\nu\nu'}
\partial_\nu (\theta^{\mu\eta} g_{\eta\nu'}(x)) .
 \label{tilde-Gamma}
\eea
The first is a consequence of the Jacobi identity resp. 
\eq{rho-g-sigma};
for the short proof see \cite{Steinacker:2008ri}.
\eq{tilde-Gamma} follows from \eq{G-def-general} and \eq{partial-theta-id}.

\paragraph{Equation of motion for test particle $\varphi$.}

The covariant e.o.m. for $\varphi$  obtained from 
the semi-classical action
\eq{covariant-action-scalar} is
\be
\Delta_{\tilde G} \varphi 
= (\tilde G^{\mu\nu}\partial_{\mu}\partial_{\nu} 
-\tilde\Gamma^\mu \partial_\mu)
\varphi = 0 \,.
\label{eom-phi-covar}
\ee
These covariant equations of motion can also be derived from the 
matrix model e.o.m.
\be
[X^a, [X^{b},\varphi]] \eta_{ab} =0
\ee
which follow from  \eq{MM-action-scalar}.
To see this, we can cast this expression in a covariant form 
as follows:
\bea
[X^a, [X^{b},\varphi]] \eta_{ab} &=& 
[X^\mu, [X^{\nu},\varphi]]\eta_{\mu\nu} 
 + [\phi^i, [\phi^j,\varphi]]\,\d_{ij} \nn\\
&=&  i[X^\mu, \theta^{\nu\eta}\partial_\eta\varphi] \eta_{\mu\nu} 
 + i[\phi^i,\theta^{\nu\eta}\partial_\nu\phi^j\partial_\eta\varphi]\,\d_{ij} \nn\\
&\sim&  -\theta^{\mu\rho}\partial_\rho
  (\theta^{\nu\eta}\partial_\eta\varphi) \eta_{\mu\nu} 
 -\theta^{\mu\rho}\partial_\mu \phi^i
 \partial_\rho(\theta^{\nu\eta}\partial_\nu\phi^j\partial_\eta\varphi) \,\d_{ij}
\nn\\
&=&  -\theta^{\mu\rho}\partial_\rho
  (\theta^{\nu\eta}\partial_\eta\varphi) \eta_{\mu\nu} 
 -\theta^{\mu\rho}\partial_\rho(
 \theta^{\nu\eta}\d g_{\mu\nu}\partial_\eta\varphi) \nn\\
&=&  -\theta^{\mu\rho}\partial_\rho
  (\theta^{\nu\eta} g_{\mu\nu}(x))\partial_\eta\varphi 
  -G^{\rho\eta}\partial_\rho\partial_\eta\varphi  \nn\\
&=& e^\sigma (\tilde \Gamma^\eta \partial_\eta\varphi 
  -\tilde G^{\rho\eta}\partial_\rho\partial_\eta\varphi)
= - e^\sigma \Delta_{\tilde G} \varphi 
\label{eom-varphi-0}
\eea
in agreement with \eq{eom-phi-covar},
using \eq{partial-theta-id}, \eq{tilde-Gamma} and denoting
\be
\d g_{\mu\nu} \equiv \partial_\mu\phi^i \partial_\nu\phi^j\,\d_{ij} .
\ee
We will see below that
in the preferred coordinates $x^\mu$ defined by the matrix model,
on-shell geometries satisfy 
\be
\tilde\Gamma^\mu \,\,\stackrel{\rm e.o.m.}{=} 0  ,
\label{gauge-fixing}
\ee
and the equations of motion for $\varphi$
simplify as 
\be
\tilde G^{\rho\eta}\partial_\rho\partial_\eta\varphi =0.
\ee
However, this non-covariant form holds only for on-shell geometries
in the preferred $x^\mu$ coordinates.

\paragraph{Equation of motion for  $X^a$.}

The same argument gives the equations of motion 
for the embedding functions $\phi^i$ in the matrix model 
\eq{YM-action-extra}, 
\be\fbox{$
\Delta_{\tilde G} \phi^i =0   $}
\label{eom-phi}
\ee
and similarly for $x^\mu \sim X^\mu$,
\be 
\Delta_{\tilde G} x^\mu = 0 .
\label{eom-X-harmonic-tree}
\ee
This is consistent with the ambiguity of the splitting  
$X^a = (X^\mu,\phi^i)$
into coordinates and scalar fields.
In particular, on-shell geometries \eq{eom-extradim}
imply harmonic coordinates, which in
General Relativity 
would be interpreted as gauge condition.

\paragraph{Equation of motion for  $\theta^{-1}_{\mu\nu}(x)$.}

Reconsider the e.o.m. for the tangential components $X^\mu$
from the matrix model \eq{YM-action-extra}:
\bea
0 = [X^b, [X^{\nu},X^{b'}]] \eta_{bb'} &=& 
[X^\mu, [X^{\nu},X^{\mu'}]] \eta_{\mu\mu'} 
 + [\phi^i, [X^{\nu},\phi^j]]\,\d_{ij} \nn\\
&=& -\theta^{\mu\rho}\partial_\rho\theta^{\nu\mu'} \eta_{\mu\mu'} 
  -\theta^{\mu\rho}\partial_\mu\phi^i
\partial_\rho(\theta^{\nu\eta}\partial_\eta \phi^j \,\d_{ij}) \nn\\
&=& -\theta^{\mu\rho}\partial_\rho(\theta^{\nu\eta} g_{\mu\eta}(x))\nn\\
&=& - e^\sigma\, \tilde \Gamma^\nu\nn\\
&=& -\theta^{\nu\nu'} G^{\rho\eta'}(x) 
  \partial_\rho\theta^{-1}_{\nu'\eta'} 
 - \theta^{\nu\eta}\theta^{\mu\rho}\partial_\rho g_{\mu\eta} 
\label{eom-extradim}
\eea
since 
$\partial_\rho \d g_{\mu\eta}(x) = \partial_\rho g_{\mu\eta}(x)$, 
i.e.
\be
 G^{\rho\eta}(x) \partial_\rho\theta^{-1}_{\eta\nu} 
= \theta^{\mu\rho}\partial_\rho g_{\mu\nu}(x).
\label{eom-extradim-theta}
\ee
As shown in \cite{Steinacker:2008ri}, 
\eq{eom-extradim-theta} together with \eq{eom-phi}
can be written in covariant form as
\be
\fbox{$
\tilde G^{\g \eta}(x)\, \tilde\nabla_\g (e^{\sigma} \theta^{-1}_{\eta\nu}) 
\, = \, e^{-\sigma}\,\tilde G_{\mu\nu}\,\theta^{\mu\g}\,\partial_\g\eta(x)$}
\label{eom-geom-covar-extra}
\ee
Here $\tilde \nabla$ denotes the Levi-Civita connection with respect to
the effective metric $\tilde G^{\mu\nu}$ \eq{G-tilde-def}.
Remarkably, this equation is also a consequence of a ``matrix'' 
Noether theorem due to a basic symmetry \eq{translations}
of the matrix model, as shown in section \ref{sec:noether}. This
means that \eq{eom-geom-covar-extra} is protected from quantum
corrections, and should be taken serious at the quantum level. 
It provides the relation between the noncommutativity 
$\theta^{\mu\nu}(x)$ and the metric $\tilde G^{\mu\nu}$.

Eq. \eq{eom-geom-covar-extra} has the structure of
 covariant Maxwell equations coupled to an external
current,
\bea
\tilde G^{\g \eta}(x)\, \tilde\nabla_\g \theta^{-1}_{\eta \nu} 
 &=& - \tilde G^{\g \eta}(x)\, \theta^{-1}_{\eta \nu}
\partial_\g \sigma + 
e^{-2\sigma}\,\tilde G_{\mu\nu} \theta^{\mu\g}\,\partial_\g\eta(x)
\label{eom-covar-2}
\eea
Therefore the noncommutativity $\theta^{\mu\nu}(x)$ 
should be completely
determined (for given boundary conditions) 
by the scalar functions $\eta(x)$ and $\rho(x)$. 
Their physical meaning remains to be elucidated.

\subsection{Frame}
\label{sec:frame}

Define 
\be
\cJ^{\eta}_\g = e^{-\sigma/2}\, \theta^{\eta\g'} g_{\g' \g}
 = - e^{-\sigma/2}\,  G^{\eta \g'} \theta^{-1}_{\g' \g} .
\ee
Then the effective metric can be written as
\be
\tilde G^{\mu\nu} 
= \cJ^{\mu}_\rho\, \cJ^{\nu}_{\rho'}\, g^{\rho\rho'}
= - (\cJ^2)^{\mu}_\rho\, g^{\rho\nu},
\ee
which is the reason for choosing the above normalization. 
Moreover, using \eq{sigma-rho-relation} we have
\be
\det \cJ = 1, \qquad 2n=4
\ee
in the case of 4 dimensions, which we assume 
in this section from now on.
Therefore $\cJ$ should be regarded as a preferred vielbein.
However,
there is no distinction between ``Lorentz'' and ``coordinate'' indices
 here, and neither local Lorentz nor general coordinate
 transformations are allowed a priori.
$\cJ^{\rho}_\nu$ satisfies the following properties
\bea 
G_{\mu\rho} \cJ^{\rho}_\nu  &=& -e^{-\sigma/2}\, \theta^{-1}_{\mu \nu}
\nn\\
(\cJ^2)^{\mu}_\rho &=& - \tilde G^{\mu \nu} g_{\nu \rho}\nn\\
tr \cJ^2 &=& - 4 e^{-\sigma} \eta \nn\\
 \cJ^{\rho}_\mu \,\tilde G_{\rho\nu}
 + \cJ^{\rho}_\nu\, \tilde G_{\mu\rho} &=& 0 \label{J-G-invar}
\eea
due to the anti-symmetry of $\theta^{-1}_{\mu\nu}$. 
Note that $\cJ^{\rho}_\nu \in so(\tilde G)$ due to \eq{J-G-invar}.

To make this more explicit,
consider (at a point $x$) a local coordinate system where
$g_{\mu\nu} = \diag(1,1,1,1)$ (in the Euclidean case)
and $\theta^{\mu\nu}$ has the form 
\be
\sqrt{\rho}\, \theta^{\mu\nu} = \left(\begin{array}{cccc} 0 & 0 & 0 & -\a \\
                                0 & 0 & -\a^{-1} & 0 \\
                                0 & \a^{-1} & 0 & 0  \\
                                \a & 0 & 0 & 0 \end{array}\right)\, ;
\label{theta-standard-general}
\ee 
this can always be achieved. 
Then locally
\bea
(\cJ^2)^{\mu}_\rho &=& -\diag(\a^2,\a^{-2},\a^{-2},\a^{2}), \nn\\
\tilde G^{\mu\nu} &=& \diag(\a^2,\a^{-2},\a^{-2},\a^2), 
\eea
and
\be
e^{-\sigma}\eta = \frac 12(\a^2+\a^{-2}).
\label{eta-alpha}
\ee
As a consequence, $\cJ$ satisfies a characteristic equation
\bea
(\cJ^2+\a^2)(\cJ^{2}+\a^{-2}) &=& 0, \nn\\
\cJ^2+ 2 e^{-\sigma}\eta  + \cJ^{-2} &=& 0
\label{chareq-J}
\eea
using \eq{eta-alpha}.
In certain cases $\cJ$ may play the role of an
almost-complex structure, which will be explored elsewhere.

\paragraph{Remarks on Minkowski case and Wick rotation.}

If $\eta_{\mu\nu}$ has Minkowski signature, there are
some rather strange modifications to the above formulae.
There are other reasons to believe
that the appropriate implementation of Minkowski signature
might require more than simply setting $\eta_{\mu\nu}=(-1,1,1,1)$.
In addition, one could 
impose $(X^0)^\dagger = - X^0$, replacing $X^0 \equiv i T$.
Then $\theta^{0i}$ would be imaginary rather than real.
There are several reasons why such a scenario might be 
preferable, but this remains to be explored.

\subsection{$SO(D)$ invariance and
normal embedding coordinates}
\label{sec:SOD}

The matrix model action is invariant under $SO(D)$ (resp. $SO(1,D-1)$)
rotations as well as translations $X^a \to X^a + c^a \one$, which 
together form the inhomogeneous Euclidean group $ISO(D)$
(resp. the Poincar\'e group). Taking advantage of this symmetry,
one can choose for any given point $p \in \cM$ adapted 
coordinates where the brane is tangential to the plane spanned by the 
first $2n$ components, i.e. $\partial_\mu \phi^i|_p =0$. 
Then the embedding metric satisfies
\be
\d g_{\mu\nu}|_p = \partial_\sigma  \d g_{\mu\nu}|_p =0 .
\label{normal-embedd-coords}
\ee
We denote such coordinates as 
``normal embedding coordinates''\footnote{They coincide with 
Riemannian normal coordinates w.r.t. the Levi-Civita connection
corresponding to $g_{\mu\nu}$, but not in general for $\tilde \nabla$
corresponding to the effective metric 
$\tilde G_{\mu\nu}$}
from now on; they exist for any given point $p \in \cM$.
Note that these are preferred matrix coordinates $x^a\sim X^a$. 

These special coordinates often simplify 
the analysis of general branes, and allow to reduce
many considerations to the case of trivially embedded branes. 
For example, it 
gives an easy way to see that the matrix model action 
in the $U(1)$ case
can be written as in \eq{S-semiclassical-general}
in terms of 
\be
4\eta(x) = \{x^a,x^b\}\{x^{a'},x^{b'}\}\eta_{aa'} \eta_{bb'}
= G^{\mu\nu}(x) g_{\mu\nu}(x).
\ee
This is so because $\eta(x)$ can be viewed as a $SO(D)$ scalar, which 
in normal embedding coordinates reduces to 
$G^{\mu\nu}(x) \eta_{\mu\nu}$. The unique covariant 
generalization is as above.

For nonabelian gauge fields, a similar argument 
strongly suggests that their effective action can be evaluated in 
normal embedding coordinates, where it should reduce to the
action computed in \cite{Steinacker:2007dq} without extra dimensions.
However, we will give an independent derivation in section
\ref{sec:nonabel}
based on the equations of motion.

\section{Moyal-Weyl plane and deformations.}
\label{sec:moyal-weyl}

A particular solution of the e.o.m.
\eq{matrix-eom} is given by the 4D Moyal-Weyl quantum plane.
Its generators $\bar X^{\mu}$  satisfy
\be
[\bar X^\mu,\bar X^\nu] = i \bar\theta^{\mu\nu} \one\, ,
\label{Moyal-Weyl}
\ee
where $\bar\theta^{\mu\nu}$
is a constant antisymmetric tensor. 
The effective geometry for the Moyal-Weyl plane is indeed
flat, given by
\bea
\bar G^{\mu\nu} &=& \bar \rho\, \bar\theta^{\mu\mu'}\,\bar\theta^{\nu\nu'} 
\bar g_{\mu'\nu'},   \nn\\
\bar\rho &=& |\bar\theta_{\mu\nu}^{-1}|^{1/2} \equiv \L_{NC}^4 .
\label{effective-metric-bar}
\eea
where the embedding metric is simply
\be
\bar g_{\mu\nu} = \eta_{\mu\nu} .
\ee
In the Minkowski case, 
we can choose coordinates where 
\be
\bar G_{\mu\nu} = \diag(-1,1,1,1),
\ee
and
\be
\sqrt{\rho}\, \theta^{\mu\nu} = \left(\begin{array}{cccc} 0 & 0 & 0 & -s\a \\
                                0 & 0 & -\a^{-1} & 0 \\
                                0 & \a^{-1} & 0 & 0  \\
                               s \a & 0 & 0 & 0 \end{array}\right)\, 
\label{theta-sign}
\ee 
where we admit $s\in\{1,i\}$. 
If we choose $s=1$, then
the original flat background metric is
\be
\bar g_{\mu\nu} = \rho^{-1}\bar \theta^{-1}_{\mu\mu'} \bar\theta^{-1}_{\nu\nu'} 
\bar G^{\mu'\nu'} 
= \diag(1,1,1,-1) \, .
\ee
On the other hand if we choose $s=i$, then $\theta^{0\mu}$ is imaginary
and the original flat background metric is
\be
\bar g_{\mu\nu} = \rho^{-1}\bar \theta^{-1}_{\mu\mu'} \bar\theta^{-1}_{\nu\nu'} 
\bar G^{\mu'\nu'} 
= \diag(-1,1,1,1) = \bar G_{\mu\nu} \, .
\ee
This illustrates the discussion in section \ref{sec:frame}.

\subsection{Linearized metric fluctuations}
\label{sec:linearized}

The Riemann and Ricci tensor for a general effective metric $\tilde G_{\mu\nu}$
are defined as usual by
\bea
\Gamma^\eta_{\mu \nu} & = & \frac 12 \tilde G^{\g \eta} 
\left( \partial_\mu \tilde G_{\nu \g} 
+ \partial_\nu \tilde G_{\g\mu} - \partial_\g \tilde G_{\mu\nu} \right),\\
R^\d_{\mu\nu \g} & = & \partial_\nu \Gamma^\d_{\mu \g} - \partial_\g\Gamma^\d_{\mu\nu} 
+\Gamma^\rho_{\mu \g} \Gamma^\d_{\rho\nu} - \Gamma^\rho_{\mu\nu} \Gamma^\d_{\rho\g},\\
R_{\mu\nu} & = & R^\d_{\mu \d\nu},\\
R & = & \tilde G^{\mu\nu} R_{\mu\nu} \, .
\eea
Now consider fluctuations of the geometry
around a flat Moyal-Weyl plane $\bar G_{\mu\nu}$,
\be
\tilde G_{\mu\nu} = \bar G_{\mu\nu} + h_{\mu\nu},
\qquad h = \bar G^{\mu\nu} h_{\mu\nu} .
\ee
Then the linearized Riemann and Ricci tensor are given by 
\bea
R^{(1)\d}_{\mu\nu\g}
 &=&  \frac12 \bar G^{\d \rho} \partial_\nu \partial_\mu  h_{\rho\g} 
- \frac12 \bar G^{\d \rho} \partial_\nu \partial_\rho  h_{\mu \g}
- \frac12 \bar G^{\d \rho} \partial_\g \partial_\mu h_{\rho\nu} + \frac12
 \bar G^{\d \rho} \bar\partial_\g \partial_\rho  h_{\mu\nu} \nn\\
R^{(1)}_{\mu\g}  &=&  
\partial^\rho \partial_{(\mu}  h_{\g)\rho} - \frac12
\partial_\mu\partial_\g  h - \frac12 \partial^\d\partial_\d h_{\mu\g} .
\eea
Now we impose the equations of motion  
in the preferred (matrix) coordinates, given by \eq{eom-extradim}
\bea
0 &=& \tilde \Gamma^\mu 
\sim \partial_\nu (\sqrt{|\tilde G_{\eta\sigma}|}\, \tilde G^{\nu\mu})
\label{gamma-eom}\\
0 &=& \Delta_{\tilde G}\, \phi^i .
\label{phi-eom}
\eea
For the fluctuations this amounts to
\be
\partial^\mu h_{\mu\nu} - \frac 12 \partial_\nu h =0 ,
\label{h-harmonic}
\ee
and the expression for the linearized Ricci tensor 
simplifies as
\be
R^{(1)}_{\mu\g}  = - \frac12 \partial^\d\partial_\d h_{\mu\g} .
\label{Ricci-harmonic}
\ee
In General Relativity, \eq{h-harmonic} would simply amount to a gauge fixing 
without dynamical content. This is not the case here: 
The preferred matrix coordinates satisfy constraints, and 
\eq{gamma-eom} resp. \eq{h-harmonic} is part of their equations of motion.

We now introduce an explicit parametrization of the 
fluctuations of  the flat Moyal-Weyl plane. 
Tangential fluctuations of the ``covariant coordinates'' can be
parametrized as 
\be
X^\mu = \bar X^{\mu} -\bar\theta^{\mu\nu} A_\nu(x) \, ,
\label{cov-coord-1}
\ee
and the $A_\nu$ can be interpreted as $U(1)$ gauge fields
on $\R^{4}_\theta$ with field strength 
$F_{\mu\nu} = \partial_\mu A_\nu - \partial_\mu A_\nu + i[A_\mu,A_\nu]$.
The variation of the Poisson tensor is easily seen to be 
$\d \theta^{-1}_{\mu\nu} = F_{\mu\nu}$. On the other hand, the variation of the
embedding metric \eq{g-def} 
\be
\d g_{\mu\nu} = \partial_\mu \d\phi^i \partial_\nu  \d\phi^j\,\d_{ij} 
\ee
is necessarily 2nd order in $\d\phi$.
Using
$\d g^{\mu\nu} = - \bar g^{\mu\mu'} \d g_{\mu'\nu'} \bar g^{\nu\nu'}$,
the combined metric fluctuation can be written as
\bea
h_{\mu\nu} 
&=& e^{\bar \sigma/2}\,\((\bar \cJ^{-1})^{\mu'}_{\nu} F_{\mu'\mu}
 + (\bar \cJ^{-1})^{\nu'}_{\mu} F_{\nu'\nu}
 - \frac 12 ((\bar \cJ^{-1})^{\rho}_{\nu} F_{\rho\sigma}\bar G^{\nu\s})\bar G_{\mu\nu}\) \nn\\
&& - (\bar \cJ^{-1})^{\mu'}_{\mu}(\bar \cJ^{-1})^{\nu'}_{\nu} \d g_{\mu'\nu'}
 + \frac 12  (\bar g^{\rho\sigma}\d g_{\rho\sigma}) \bar G_{\mu\nu} 
\label{hmunu-general}
\eea
since $e^\sigma = \sqrt{|g_{\mu\nu}|/|\theta^{-1}_{\mu\nu}|}$
in 4 dimensions \eq{sigma-rho-relation}.
As a check, we note that $h = \bar g^{\rho\sigma}\d g_{\rho\sigma}$
depends only on the embedding part, cf. \eq{G-g-4D}.

\subsection{$U(1)$ fluctuations and gravitational waves}

For $\d g_{\mu\nu}=0$,
the on-shell $U(1)$ fluctuations  around flat Moyal-Weyl space
satisfy the Maxwell equations 
\be
\partial^\mu F_{\mu\nu} =0
\label{maxwell}
\ee
which implies as usual
\be
\partial^\rho\partial_\rho F_{\mu\nu} =0 .
\label{box-F}
\ee
Together with \eq{Ricci-harmonic} and \eq{hmunu-general}, it follows that
these $U(1)$ metric fluctuations are Ricci-flat,
\be
R_{\mu\nu}  =0 
\label{ricci-flat}
\ee
which was first observed by Rivelles \cite{Rivelles:2002ez}.
This can  also be seen as a consequence of 
$\Delta_{\tilde G} X^\mu =0$.
We verify below that the corresponding Riemann tensor is 
indeed non-trivial, thus obtaining a
parametrization of the 2 on-shell degrees of freedom of
gravitational waves through $U(1)$ gauge fields.
Note also that \eq{ricci-flat} is consistent with the
Einstein-Hilbert action induced upon quantization \cite{Steinacker:2007dq}.

\paragraph{Linearized Riemann tensor.}

Consider $U(1)$ fluctuations given by  plane waves
\be
A_\mu(x) = A_\mu e^{ikx} 
\ee
such that $F_{\mu\nu} = i(k_\mu A_\nu - k_\nu A_\mu) \neq 0$.
Then
\bea
h_{\mu\nu} &=&  i(\tilde k_{\nu}A_{\mu} - k_{\mu}\tilde A_{\nu})
+i(\tilde k_{\mu}A_{\nu} - k_{\nu}\tilde A_{\mu})
- \frac 12 i\bar G_{\mu\nu}\,
((\tilde k_{\rho}A_{\sigma} -  k_{\rho}\tilde A_{\sigma})\bar G^{\rho\sigma}) \nn\\
&=& +i(\tilde k_{\mu}A_{\nu}+\tilde k_{\nu}A_{\mu})
 - i (k_{\mu}\tilde A_{\nu} + k_{\nu}\tilde A_{\mu}) 
+ ie^{\bar \sigma/2}\,
\bar G_{\mu\nu}\,(\bar\theta^{\rho\sigma}k_{\rho}A_{\sigma})
\label{h-abdown}
\eea
where
\be
\tilde k_\mu = e^{\bar \sigma/2}\,(\cJ^{-1})^\rho_\mu k_\rho ,
\ee
noting that
\be
\tilde k_{\rho}A_{\sigma} \bar G^{\rho\sigma} = 
- e^{\bar \sigma/2}\,\bar\theta^{\rho\sigma}k_{\rho}A_{\sigma}
= k_{\rho}\tilde A_{\sigma} \bar G^{\rho\sigma} .
\ee
Observe that the term 
$i (k_{\mu}\tilde A_{\nu} + k_{\nu}\tilde A_{\mu}) $
has the form of a diffeomorphism generated by $\tilde A_{\nu}$, and
therefore does not contribute to the Riemann tensor. 
In particular, the deformation is 
conformally flat if $\tilde k_{\mu} \sim k_\mu$, which 
means that $k_\mu$ is an eigenvector of $\cJ$. 
Using \eq{h-abdown} for the metric fluctuations gives
\bea
R^{(1)}_{\sigma\mu\nu\g}
&=& \bar G_{\d \sigma}\, R^{(1)\d}_{\mu\nu\g} 
=  \frac12 \( \partial_\nu \partial_\mu h_{\sigma\g} 
-  \partial_\nu \partial_\sigma h_{\mu \g}
- (\g \leftrightarrow \nu)\) \nn\\
&=& - \frac i2 \Big(
( k_\mu \tilde k_{\s} - \tilde k_{\mu} k_\s)
(k_\nu A_{\g} - A_{\nu}k_\g)
+ (k_\mu A_{\s}  - k_\s A_{\mu}) 
(k_\nu\tilde k_{\g} - \tilde k_{\nu}k_\g)\Big) \nn\\
&&  - \frac i2e^{\bar \sigma/2}\,( k_\nu k_\mu  \bar G_{\s\g} 
  - k_\nu k_\s  \bar G_{\mu\g} 
- k_\g k_\mu  \bar G_{\s\nu} 
+ k_\g k_\s  \bar G_{\mu\nu}) 
(\bar\theta^{\rho\sigma}k_{\rho}A_{\sigma}) .
\label{Riemann-first}
\eea
This implies in particular
\bea
R^{(1)}_{\sigma\mu\nu\g} \bar\theta^{\nu\g}
&=& - i (k_\mu A_{\s}  - k_\s A_{\mu}) 
 (k_\nu\tilde k_{\g}\bar\theta^{\nu\g}) 
\, = \, F_{\s\mu}\,(k_\nu\tilde k_{\g}\bar\theta^{\nu\g}) , 
\eea
which vanishes\footnote{note that $A \not\sim k$ for 
nontrivial $h_{\mu\nu}$} 
only if $k_\nu\tilde k_{\g}\bar\theta^{\nu\g}=0$. 
In that case, let $A_\mu = \tilde k_\mu$ 
and \eq{Riemann-first} gives 
\bea
R^{(1)}_{\sigma\mu\nu\g}
&=& - i (k_\mu \tilde k_{\s} - \tilde k_{\mu} k_\s)
(k_\nu \tilde k_{\g} - \tilde k_{\nu}k_\g) \neq 0
\qquad \mbox{provided}\quad k \not\sim \tilde k .
\eea
Now it is easy to see that $\cJ$ has no real eigenvector in the Euclidean case
(since  $\cJ$ consists essentially of 2 rescaled complex 
structures as discussed in section \ref{sec:frame}),
hence
\be
R^{(1)}_{\sigma\mu\nu\g} \neq 0 .
\ee
The same is true in the Minkowski case, provided
we use imaginary $\theta^{0\mu}$ as discussed in section \ref{sec:frame};
this seems to be the better choice.
Thus the $U(1)$ metric fluctuations have 2
non-trivial physical degrees of freedom, and therefore describe the
physical gravitons on flat $\R^4$. 
If we insist on real $\theta^{0\mu}$ in the Minkowski case, then
there would be a preferred direction $k \sim\tilde k$ where only
one graviton polarization survives.

\subsection{Fluctuations of the embedding}

Now consider on-shell fluctuations of the embedding metric. 
The linearized Ricci tensor for $F_{\mu\nu} =0$ is 
\be
R^{(1)}_{\mu\nu}  =  \frac12 \Delta_{\bar G} \(
 (\bar \cJ^{-1})^{\mu'}_{\mu}(\bar \cJ^{-1})^{\nu'}_{\nu} \d g_{\mu'\nu'}
 - \frac 12  (\bar g^{\rho\sigma}\d g_{\rho\sigma}) \bar G_{\mu\nu} \)
\ee
using \eq{Ricci-harmonic},
which vanishes for
\be
\Delta_{\bar G} \d g_{\mu\nu} =0 .
\ee
This is a rather restrictive condition, which however does have
nontrivial solutions, in particular the plane wave solution
\be
\d\phi^i = p^i\, e^{i k x} ,
\label{planewave-embedd}
\ee 
so that
$\partial_\mu \phi^i = k_\mu\,\phi^i$. Imposing the e.o.m. \eq{eom-phi-covar} 
leads to light-like momentum 
$k_\mu k_\nu \bar G^{\mu\nu} =0$, and 
$R^{(1)}_{\mu\nu}  =0$ follows. 
While such plane-wave deformations correspond to trivial (flat) 
deformations of the embedding metric $g_{\mu\nu}$, they
do lead to non-trivial deformations of the 
effective metric $\tilde G_{\mu\nu}$, as verified below by computing
the corresponding Riemann tensor.

This result holds for every component $\phi^i$, and it may appear 
at first that there are too many ``gravitons''. 
However, recall that the embedding fluctuations 
only contribute at second-order in $\d\phi$. Therefore these plane
waves cannot simply be superimposed, and 
should presumably {\em not} be interpreted as gravitational 
waves. This suggests that these Ricci-flat embedding fluctuations 
should be interpreted as gravitational field due to some  
nontrivial mass distributions;
this is expected to happen e.g. for the analog of the
Schwarzschild geometry.
Moreover, one should keep in mind that only 
certain combinations of these embedding and $U(1)$ fluctuations may
survive at the quantum level; this is suggested by the fact that
only the ``tangential'' e.o.m. are 
ensured by Noether's theorem \eq{cons-tang-explicit-1}.
It would be very important to understand better 
the relation between the $U(1)$ and the embedding fluctuations.

\paragraph{Linearized Riemann tensor.}

For $F_{\mu\nu} =0$, the linearized Riemann tensor is
\bea
R^{(1)}_{\sigma\mu\nu\g}
&=&  - \frac12 \Big( (\cJ^{-1})_\g^{\g'} (\cJ^{-1})_\sigma^{\sigma'} 
\partial_\nu \partial_\mu \d g_{\sigma'\g'}
-  (\cJ^{-1})_\g^{\g'}  (\cJ^{-1})_\mu^{\mu'}
\partial_\nu \partial_\sigma \d g_{\mu'\g'}- (\g \leftrightarrow\nu)\Big) \nn\\
&& +\frac 14\(\bar G_{\sigma \g}\partial_\nu \partial_\mu  h
-\bar G_{\sigma \nu}\partial_\g \partial_\mu  h
-\bar G_{\mu \g}\partial_\nu \partial_\sigma  h  
+ \bar G_{\mu \nu}\partial_\g \partial_\sigma  h \) 
\label{Riemann-embedd}
\eea
where $h = \bar g^{\mu\nu}\d g_{\mu\nu}$.
For plane wave deformations \eq{planewave-embedd}, this is
\bea
R^{(1)}_{\sigma\mu\nu\g}
&=& \frac12 e^{-\sigma}\, (\tilde k_\sigma k_\mu - \tilde k_\mu k_\sigma)
(\tilde k_\nu k_\g - \tilde k_\g k_\nu) (p^i p^j \d_{ij})  \nn\\
&& - \frac 14\Big(\bar G_{\sigma \g}k_\nu k_\mu 
-\bar G_{\sigma \nu}k_\g k_\mu 
-\bar G_{\mu \g}k_\nu k_\sigma
+ \bar G_{\mu \nu}k_\g k_\sigma \Big)h 
\label{Riemann-embedd-wave}
\eea
which is indeed nontrivial for $k \neq \tilde k$.

\section{Noether theorem and semi-classical conservation laws}
\label{sec:noether}

The basic matrix model \eq{YM-action-0} enjoys the 
translational symmetry
\be
X^a \to X^a + c^a\one
\label{translations}
\ee
As shown in \cite{AbouZeid:2001up,Steinacker:2008ri}, 
this symmetry leads to the conservation law 
\be
[X^a,T^{a'c}] \eta_{aa'} =0 
\label{e-m-cons}
\ee
which was also discussed in \cite{Das:2002jd} in a 
somewhat different context, cf. \cite{Okawa:2001if}.
It can also be verified directly using
the matrix e.o.m. \eq{matrix-eom}.
Here
\be
T^{ab} =  [X^a,X^c][X^b,X^{c'}]\eta_{cc'} 
  + [X^b,X^c][X^a,X^{c'}] \eta_{cc'}
  - \frac 12 \eta^{ab} [X^d,X^c][X^{d'},X^{c'}] \eta_{dd'} \eta_{cc'}
\label{T-ab}
\ee
is the matrix - ``energy-momentum tensor''.
However, its interpretation and consequences are 
quite surprising in the present context, as we will show.
Note first that the indices run from $1$ to $D$.
In the semi-classical 
or geometrical limit, $T^{ab}$
can be written in terms of the geometrical
quantities,
\bea
T^{\mu\nu} &=& [X^\mu,X^\rho][X^\nu,X^{\rho'}]\eta_{\rho\rho'} 
  + [X^\nu,X^\rho][X^\mu,X^{\rho'}] \eta_{\rho\rho'}
  - \frac 12 \eta^{\mu\nu} [X^\rho,X^\sigma][X^{\rho'},X^{\sigma'}] 
 \eta_{\rho\rho'} \eta_{\sigma\sigma'} \nn\\
 && + [X^\mu,\phi_r][X^\nu,\phi_r]
 + [X^\nu,\phi_r][X^\mu,\phi_r] 
  - \eta^{\mu\nu} [X^\rho,\phi_r][X^{\rho'},\phi_r] \eta_{\rho\rho'} 
 - \frac 12 \eta^{\mu\nu} [\phi_r,\phi_s][\phi_r,\phi_s]  \nn\\
&\sim&- 2  G^{\mu\nu} + 2 \eta^{\mu\nu} \eta(y) \nn\\
T^{\mu r} &=& 
 [X^\mu,X^\rho][\phi_r,X^{\rho'}]\eta_{\rho\rho'} 
  + [\phi_r,X^\rho][X^\mu,X^{\rho'}] \eta_{\rho\rho'} 
 + [X^\mu,\phi_t][\phi_r,\phi_{t'}]\eta_{tt'} 
 + [\phi_r,\phi_{t'}] [X^\mu,\phi_t] \eta_{tt'} \nn\\
 &\sim& - 2\theta^{\rho\rho'} \theta^{\mu\mu'} g_{\rho'\mu'}(y)
 \partial_{\rho}\phi_r 
 = -2 G^{\rho\nu}\partial_{\rho}\phi_r  \nn\\
T^{r s} &=&[\phi_r,X^\rho][\phi_s,X^{\rho'}]\eta_{\rho\rho'} 
  + [\phi_s,X^\rho][\phi_r,X^{\rho'}] \eta_{\rho\rho'}
+ [\phi_r,\phi_t][\phi_s,\phi_{t'}]\eta_{tt'} 
+ [\phi_s,\phi_{t'}][\phi_r,\phi_{t'}] \eta_{tt'} \nn\\
&&  - \frac 12 g^{r s} [X^\rho,X^\sigma][X^{\rho'},X^{\sigma'}] 
 \eta_{\rho\rho'} \eta_{\sigma\sigma'} \nn\\
 &\sim& -2 G^{\mu\nu}\partial_\mu \phi_r \partial_\nu \phi_s
    + 2 \eta^{r s} \eta(y) 
\eea
where
\bea
g_{\mu\nu}(y) &=& \eta_{\mu\nu} + \partial_{\mu} \phi_r\partial_{\nu}
\phi_r = \eta_{\mu\nu} + \d g_{\mu\nu} , \nn\\
G^{\mu\nu}(y) &=& \theta^{\mu\mu'}\theta^{\nu\nu'}g_{\mu\nu}(y)\nn\\
{\bf \eta}(y) &=&
\frac 14\,  G^{\mu\nu} g_{\mu\nu}(y)
= \frac 14\,g_{\rho\sigma}(y) 
\theta^{\rho\rho'}\theta^{\sigma\sigma'} g_{\rho'\sigma'}(y) 
\eea
as introduced before.
We want to elaborate the semi-classical meaning of
the conservation laws 
\bea
&& [X^{\mu'},T^{\mu \nu}] \eta_{\mu'\mu} 
 + [\phi_r,T^{r \nu}]  = 0 \label{cons-tang}\\
&&  [X^{\mu'},T^{\mu s}] \bar \eta_{\mu'\mu} 
 + [\phi_r,T^{r s}]  = 0 \label{cons-scalar},
\eea
which we will refer to as tangential resp. scalar conservation law.
Using
\bea
i\frac 12 [\phi_r,T^{r \nu}] 
&=& \theta^{\mu\eta}\partial_\mu \phi_r 
 \partial_\eta(G^{\nu\rho} \partial_{\rho}\phi_r) 
= \theta^{\mu\eta}\partial_\eta(G^{\nu\rho}\d g_{\mu\rho}) , \nn\\ 
-i [X^{\nu},T^{\mu s}] \bar \eta_{\nu\mu} 
&=& \theta^{\nu\sigma} \partial_\sigma( - 2\theta^{\rho\rho'} \theta^{\mu\mu'} \eta_{\rho'\mu'}
 \partial_{\rho}\phi_s) \eta_{\nu\mu} , \nn\\
-i [\phi_r,T^{r s}]
&=& -2\theta^{\eta\sigma}\partial_\sigma
(G^{\mu\nu}\d g_{\eta\mu} \partial_\nu \phi_s )
+ 2\theta^{\eta\sigma}\partial_\eta \phi_s \partial_\sigma \eta(y) 
\eea
the tangential conservation law \eq{cons-tang} gives
\bea
0 &=& -i[X^{\mu'},T^{\mu \nu}]\bar  g_{\mu'\mu} 
 -i [\phi_r,T^{r \nu}] \nn\\
&=& \theta^{\mu'\rho}\partial_\rho
(- 2  G^{\mu\nu} + 2\bar  g^{\mu\nu} \eta(y))\bar  g_{\mu'\mu}
 -2 \theta^{\mu\eta}\partial_\eta(G^{\nu\rho}\d g_{\mu\rho}) \nn\\
&=& 2 \theta^{\nu\rho}\partial_\rho\eta(y)
 -2 \theta^{\mu\eta}\partial_\eta(G^{\nu\rho} g_{\mu\rho}(y))  
\label{cons-tang-explicit-1}
\eea 
which using $\partial_\eta\theta^{\eta\mu} =
\theta^{\mu\eta}\rho^{-1}\partial_\eta\rho$ \eq{partial-theta-id}
can be written as
\be \fbox{$
\partial_\eta(\rho\hat\theta^{\eta\nu})
= \rho\,\theta^{\rho\nu}\partial_\rho\eta(y) $}
\label{conservation-general-1}
\ee
introducing the antisymmetric (!) matrix
\be
\hat \theta^{\nu\eta} 
=  G^{\nu\rho} g_{\rho\mu}(y)\theta^{\mu\eta} 
= -  G^{\nu\rho} \theta^{-1}_{\rho\mu} G^{\mu\eta} 
=- \theta^{\nu\mu'}g_{\mu'\eta'}\theta^{\eta'\eta} g_{\eta\nu'}\theta^{\nu'\eta} 
= - e^\sigma {(\cJ^2)^\nu}_\mu \theta^{\mu\eta} 
\label{theta-hat}
\ee
which we will encounter again. 
Similarly, the scalar conservation law \eq{cons-scalar} gives
\bea
-i [X^{\nu},T^{\mu s}] \eta_{\nu\mu} 
-i [\phi_r,T^{r s}]
&=& - 2 \theta^{\eta\sigma} \partial_\sigma
(G^{\rho\mu} \partial_{\rho}\phi_s) \eta_{\eta\mu} \nn\\
&& -2\theta^{\eta\sigma}\partial_\sigma
(G^{\rho\mu}\d g_{\eta\mu} \partial_\rho \phi_s ) 
+ 2\theta^{\eta\sigma}\partial_\eta \phi_s \partial_\sigma \eta(y) \nn\\
&=& - 2 \theta^{\eta\sigma} G^{\mu\rho} g_{\eta\mu}(y)\partial_\sigma
 \partial_{\rho}\phi_s \nn\\
&& - 2 \theta^{\eta\sigma} \partial_\sigma
(G^{\mu\rho} g_{\eta\mu}(y)) \partial_{\rho}\phi_s 
+ 2\theta^{\rho\sigma} \partial_\sigma \eta(y)\partial_\rho \phi_s\nn\\
&=& - 2 \theta^{\eta\sigma} g_{\eta\mu}(y) G^{\mu\rho} \partial_\sigma
 \partial_{\rho}\phi_s 
= - 2 \hat\theta^{\rho\sigma} \partial_\sigma
 \partial_{\rho}\phi_s =0 \nn
\eea
using the tangential conservation law 
\eq{cons-tang-explicit-1} in the last step, 
and noting that $\hat \theta^{\mu\nu}$ is antisymmetric.
Hence the scalar conservation law
is a consequence of the tangential one. This 
is somewhat surprising, but reflects the 
$2n$-dimensional nature of the sub-manifold $\cM \subset \R^D$.

The tangential conservation law \eq{conservation-general-1}
can be rewritten using $|\tilde G_{\mu\nu}|^{1/2}= \rho e^{\sigma}$ 
\eq{rho-g-sigma} and $\partial_\eta(\rho\hat\theta^{\eta\nu})
= \sqrt{|\tilde G|}\, \tilde\nabla_\mu (e^{-\sigma}\hat \theta^{\mu\nu})$
as follows:
\be
\fbox{$
\tilde\nabla_\mu (e^{-\sigma}\hat \theta^{\mu\nu})
= e^{-\sigma}\theta^{\mu\nu}\partial_\mu\eta(y) .$}
\label{conservation-general}
\ee
Using \eq{theta-hat}, this coincides precisely with
the e.o.m. \eq{eom-geom-covar-extra} 
$$
\tilde\nabla^\mu (e^{\sigma} \theta^{-1}_{\mu\nu})   
= e^{-\sigma} \tilde G_{\nu\nu'}\theta^{\nu'\mu}\partial_\mu\eta(y) .
$$
It is easy to verify the integrability condition for this conservation 
law
\bea
\tilde\nabla_\nu\tilde\nabla_\mu (e^{-\sigma}\hat \theta^{\mu\nu})
&=& \tilde\nabla_\nu(e^{-\sigma}\theta^{\mu\nu}\partial_\mu\eta(y)) .
\eea
Since the above derivation is based on an
exact conservation law of the underlying matrix model,
\eq{conservation-general} can be trusted also at 
the quantum level. 
Indeed the integration measure for the matrix model
quite trivially respects the translational
symmetry, hence no ``anomalies'' are expected.

\paragraph{Some results for the embedding metric $g_{\mu\nu}$.}

The above conservation law \eq{conservation-general} implies also 
some interesting results for the embedding metric, notably
\bea
\tilde\nabla^\rho g_{\rho\mu}
&=&  2\partial_\mu (e^{-\sigma}\eta)  \label{g-conservation}\\
\theta^{\mu\rho} \tilde\nabla_\rho (e^{\sigma} g_{\mu\eta})
&=& \tilde G_{\nu\nu'}\theta^{\nu'\mu}\partial_\mu\eta(y)
\label{g-theta} \\
(\Delta_{\tilde G} x^a)\, \partial_\nu x^b \eta_{ab} 
&=& 0 .  \label{phi-harmonic-2nd}
\eea
The last relation becomes
more transparent in  normal embedding coordinates, where it 
reduces to $\Delta_{\tilde G} x^\mu=0$.
This means that $ \Delta_{\tilde G} \vec x$ is normal to 
$T\cM$ using the metric $\eta_{ab}$.
This is somewhat weaker than (but a consequence of) 
the  ``bare'' matrix model equations
$\Delta_{\tilde G} \phi^i = \tilde\Gamma^\mu =0$, 
which would imply harmonic embeddings for all coordinates 
$x^a$; see also Appendix B of \cite{Steinacker:2008ri}.

\begin{description}
\item[proof of \eq{g-conservation}]:
 To see this, 
consider the following identity:
\bea
\partial_\rho\eta &=& \frac 14 \tilde G^{\mu\nu}
\tilde\nabla_\rho(e^{\sigma}g_{\mu\nu})  \nn\\
&=& \frac 12 \tilde\nabla_\rho \theta^{\mu\nu}
   (g\theta g)_{\mu\nu}
 + \frac 12 \tilde\nabla_\rho g_{\mu\nu} G^{\mu\nu} \nn\\
&=& \frac 12 \tilde\nabla_\rho \theta^{\mu\nu}
   (g\theta g)_{\mu\nu}
 + 2 \partial_\rho\eta
- 2 \eta e^{-\sigma}\partial_\rho e^{\sigma}
\eea
which implies
\bea
 (g\theta g)_{\a\b} \tilde\nabla_\mu \theta^{\a\b}
&=& -2\partial_\mu\eta + 4 \eta e^{-\sigma}\partial_\mu e^{\sigma}
= -2e^{2\sigma}\partial_\mu(\eta e^{-2\sigma}) \\
2e^{2\sigma}\theta^{\nu\mu}\partial_\mu(\eta e^{-2\sigma})
&=& -(g\theta g)_{\a\b} \theta^{\nu\mu}\tilde\nabla_\mu \theta^{\a\b}  
= -2(g G)_\a^{\mu}\tilde\nabla_\mu \theta^{\nu\a}\\
\theta^{\nu\mu}\partial_\mu\eta 
&=& (g G)_\a^{\mu}\tilde\nabla_\mu \theta^{\a\nu}
+\eta\theta^{\nu\mu} e^{-2\sigma}\partial_\mu e^{2\sigma}\nn\\
&=& e^{\sigma}\tilde\nabla_\mu (e^{-\sigma}\hat\theta^{\mu\nu})
+ \theta^{\nu\a} G^{\mu\rho}\tilde\nabla_\mu g_{\rho\a}
+2\eta\theta^{\nu\mu} e^{-\sigma}\partial_\mu e^{\sigma}
\eea
using the Jacobi identity in the 2nd line.
This identity together with 
the on-shell conservation law \eq{conservation-general} 
implies \eq{g-conservation}.

\item[proof of \eq{phi-harmonic-2nd}]:
this follows simply from \eq{eom-geom-covar-extra} together with
\be
 e^{-\sigma}\theta^{\mu\rho} \tilde\nabla_\rho (e^\sigma g_{\mu\eta})
 = -\tilde\nabla_\rho (\theta^{\rho\mu} g_{\mu\eta}) 
 = \tilde\nabla_\rho (G^{\rho\mu} \theta^{-1}_{\mu\eta}) 
 = \tilde G^{\rho\mu} \tilde\nabla_\rho (e^\sigma\theta^{-1}_{\mu\eta}) .
\ee

\item[proof of \eq{phi-harmonic-2nd}]:
consider
\bea
\partial^\mu g_{\mu\nu} &=&
 \tilde G^{\mu\rho}\partial_\mu\partial_\rho x^a \partial_\nu x^b \eta_{ab} 
+ \tilde G^{\mu\rho}\partial_\mu x^a\partial_\rho
\partial_\nu x^b \eta_{ab}\nn\\
&\sim& (\Delta_{\tilde G} x^a + \tilde \Gamma^\rho\partial_\rho x^a)
\partial_\nu x^b \eta_{ab} 
+\frac 12 \tilde G^{\mu\rho}\partial_\nu g_{\mu\rho}\nn\\
&=& \Delta_{\tilde G} x^a\, \partial_\nu x^b \eta_{ab} 
+ \tilde \Gamma^\rho  g_{\rho\nu}
+2\partial_\nu (e^{-\sigma}\eta) 
- \frac 12 (\partial_\nu \tilde G^{\mu\rho}) g_{\mu\rho} 
\label{g-cons-2}
\eea
(where $x^a$ are treated as scalar fields for $a,b=1,...,D$).
Now for any symmetric tensor $g_{\mu\nu}$, one has
\bea
\tilde\nabla^\mu g_{\mu\nu} &=& \partial^\mu g_{\mu\nu} 
- \tilde\Gamma^\sigma g_{\sigma\nu}
+ \tilde G^{\mu\rho}\tilde\Gamma_{\rho\nu}^\sigma g_{\mu\sigma} \nn\\
&=&\partial^\mu g_{\mu\nu} 
- \tilde\Gamma^\sigma g_{\sigma\nu}
+ \frac 12 g_{\mu\rho} \partial_\nu \tilde G^{\mu\rho} 
\eea
hence
\be
\tilde\nabla^\mu g_{\mu\nu} = \Delta_{\tilde G} x^a\, \partial_\nu x^b \eta_{ab} 
 +2\partial_\nu (e^{-\sigma}\eta) 
\ee
which together with \eq{g-conservation} gives \eq{phi-harmonic-2nd}.

\end{description}

\section{Nonabelian gauge fields}
\label{sec:nonabel}

To minimize notational conflicts, we denote the basic dynamical matrices 
with $Y^a$ in this section. 
Consider the same matrix model as above
\be
S_{YM} = - Tr [Y^a,Y^b] [Y^{a'},Y^{b'}] 
\eta_{aa'}\eta_{bb'}
\label{YM-MM-nonabel}
\ee
but for  matrix backgrounds of the form
\be
Y^a = \left(\begin{array}{l}Y^\mu \\ Y^i \end{array}\right) = 
\left\{\begin{array}{ll} X^\mu\otimes \one_n,  & \quad a=\mu = 1,2,..., 2n,   \\ 
\phi^i \otimes \one_n, & 
\quad a = 2n+i, \,\, i = 1, ..., D-2n . \end{array} \right.
\ee
We want to understand general fluctuations around the above background.
Since the $U(1)$ components describe the geometry, 
we expect to find $su(n)$-valued gauge fields as well as
 scalar fields in the adjoint.
It turns out that the following gives an appropriate parametrization 
of these general fluctuations:
\be
\left(\begin{array}{l}Y^\mu \\ Y^i \end{array}\right) = 
\left(\begin{array}{l} X^\mu\otimes \one_n + \cA^\mu   \\ 
\phi^i\otimes \one_n  + \Phi^i
+ \cA^\rho \partial_\rho (\phi^i\otimes \one_n  + \Phi^i)\end{array} \right)
\sim (1+ \cA^\nu\partial_\nu)\left(\begin{array}{l} X^\mu\otimes \one_n  \\ 
\phi^i\otimes \one_n  + \Phi^i\end{array} \right)
\label{nonabelian-SW}
\ee
where
\bea
\cA^\mu &=&  \cA^\mu_\a \otimes \lambda^\a
\,\, = \,\, - \theta^{\mu\nu} A_{\nu,\a} \otimes \lambda^\a, \nn\\
\Phi^i  &=&  \Phi^i_\a\otimes \l^\a 
\eea
parametrize the $su(n)$-valued gauge fields resp. scalar fields, and
$\lambda^\a$ denotes the generators of $su(n)$.
This amounts to the 
leading term in a Seiberg-Witten (SW) map 
\cite{Seiberg:1999vs}, which relates
noncommutative and commutative gauge theories with the appropriate 
gauge transformations\footnote{We could drop the term 
$\cA^\rho \partial_\rho \Phi^i$ for the nonabelian scalars
at leading order
as long as they do not acquire a VEV (in which case they would
contribute to the geometry), 
however the term $\cA^\rho \partial_\rho \phi^i$ 
does contribute to the effective action for $A_\mu$.}. 
This SW parametrization can be characterized by requiring that the 
noncommutative gauge transformation 
$\d_{nc} Y^a = i [Y^a,\Lambda]$ 
induces for the $su(n)$ components
$A_\mu = A_{\mu,\a} \la^\a$ and $\Phi^i$
the ordinary gauge transformations 
\bea
\d_{cl} A_\mu &=& i[A_\mu,\Lambda]_{su(n)} \, +\, \partial_\mu\Lambda(x) 
\quad + O(\theta)\nn\\
\d_{cl} \Phi^i &=& 
i [\Phi^i,\Lambda]_{su(n)} \quad + O(\theta^2) ,
\eea 
cf. \cite{Jurco:2000fb}.
Here the subscript $[\Phi^i,\Lambda]_{su(n)}$
indicates that only the commutator of the explicit $su(n)$ generators
is to be taken, but not the $O(\theta)$ contributions from the 
Poisson bracket. This strongly suggests that the matrix model action
expressed in terms of these $A_\mu$ should reduce to a conventional
gauge theory in the semi-classical limit;
 this was verified in \cite{Steinacker:2007dq}.

In our context, the SW map \eq{nonabelian-SW} can be understood geometrically
in a very simple way. Recall that the $U(1)$ sector  
(i.e. the components
proportional to $\one_n$) describes the
geometrical degrees of freedom:  
in the geometrical limit,
$X^a = (X^\mu,\phi^i)$ become functions
\be
x^a = (x^\mu,\phi^i(x))
\ee
on $\cM$ which describe the embedding of the $2n$-dimensional brane
$\cM \subset \R^D$. Then
the one-form $A_\mu dx^\mu$ together with the 
Poisson tensor determines a tangential vector field
\be
A_\mu e^\mu = A_\mu\theta^{\mu\nu}\partial_\nu = \cA^\nu \partial_\nu
\quad \in T_p\cM ,
\ee 
whose push-forward in the ambient space $\R^D$ 
\be
\cA^\nu\partial_\nu x^a
\cong \cA^\nu(\d_\nu^\mu,\partial_\nu \phi^i)
\ee
coincides with the fluctuations $\d X^a = Y^a - X^a$ 
of the dynamical matrices in \eq{nonabelian-SW} 
(for vanishing $\Phi^i$).
This provides the link between gauge fields 
and ``covariant coordinates'' $X^a$.
The nonabelian scalar fields $\Phi^i$
can be thought of as coordinate functions
 of $\cM$ embedded in further 
extra dimensions corresponding to $su(n)$. They
behave as scalar fields, but might contribute to the
background geometry if they acquire a non-trivial VEV,
completely analogous to the $U(1)$ components $\phi^i$.

After this preparation, 
we claim that the effective action for $su(n)$-valued
gauge fields $A_{\mu}$ 
on general $\cM_\theta^{2n}\subset \R^D$
in the matrix model \eq{YM-MM-nonabel}
in the semi-classical limit is
\bea
S_{YM}[\cA]  &=& Tr \Big( G^{\mu\mu'} G^{\nu\nu'} F_{\mu\nu}\, F_{\mu'\nu'}
 -  F_{\mu'\nu'} F_{\mu\nu} \hat\theta^{\mu'\nu'} \theta^{\mu\nu} 
- 2 F_{\mu'\mu} F_{\nu'\nu}\hat\theta^{\mu'\nu'}\theta^{\nu\mu}\nn\\  
&& \qquad + \frac 12 \eta\theta^{\mu\nu}\theta^{\mu'\nu'}
\big(F_{\mu\nu}F_{\mu' \nu'} + 2 F_{\mu\nu'} F_{\nu\mu'} \big)  \Big) \nn\\
&\sim&  \int d^{2n} x\, |\tilde G_{\mu\nu}|^{1/2}
e^{\sigma} \, \tilde G^{\mu\mu'} \tilde G^{\nu\nu'}
\tr(F_{\mu\nu}\,F_{\mu'\nu'})\,\, 
-\,\, S_{NC} 
\label{action-expanded-2}
\eea
where $F_{\mu\nu} = \partial_\mu A_\nu - \partial_\nu A_\mu + i
[A_\mu,A_\nu]$
is the $su(n)$-valued field strength, 
and
\bea
S_{NC} &=& S_{NC,1} + S_{NC,2} \nn\\
&=&  \int d^{2n} x\,\rho\, \tr \((F\theta) (F \hat\theta) - 2 (F \theta F\hat\theta)\)
- \frac 12\int  d^{2n} x\, \eta\rho\, 
\tr \Big((F\theta)(F \theta) - 2 (F \theta F\theta))\Big) \nn\\
&\stackrel{n=2}{=}& - 2  \int  \eta(x) \, \tr F\wedge F  \; .
\label{S-NC-def}
\eea
Note that the embedding 
metric $g$ enters the ``would-be topological term''
$S_{NC}$
only through $\eta$ resp. $\hat \theta^{\mu\nu}$.
The contraction of the tensor indices is spelled out in \eq{action-expanded-2}.
The last line holds for 4-dimensional branes, and can be seen using
\be
\frac 12 (F\wedge F)_{\mu\nu\rho\s} \hat\theta^{\mu\nu}\theta^{\rho\s}
= (F_{\mu\nu}\hat\theta^{\mu\nu}) (F_{\rho\s}\theta^{\rho\s})
 +  2 F_{\mu\s} F_{\nu\rho} \hat\theta^{\mu\nu}\theta^{\rho\s} 
\label{FF-theta}
\ee
and 
$\hat\theta\wedge\theta = \eta(y) \theta\wedge\theta$, see
\cite{Steinacker:2007dq}.
For the 4-dimensional matrix model this was first obtained
through a direct but rather non-transparent
computation of the action \cite{Steinacker:2007dq}, 
requiring the 2nd order Seiberg Witten map.
We will show below that \eq{action-expanded-2} holds also on general
non-trivially embedded $2n$ -dimensional NC branes $\cM \subset \R^D$,
by showing that the corresponding equations of motion coincide with
the equation \eq{cons-nonabelian} derived from the 
covariant conservation law \eq{cons-tang}, which 
in turn follow from the basic matrix equations 
of motion \eq{matrix-eom}.

\paragraph{Nonabelian scalars.}

It follows easily along the lines of section \ref{sec:extradim} that
the geometrical action for the $su(n)$ -valued scalars 
$\Phi^i_\a$ in the matrix model at leading order is
\bea
S_{YM}[\Phi] &\sim&  2\int d^4 x\, \sqrt{|\tilde G|}\,
\tilde G^{\mu\nu} \tr (D_\mu \Phi^i  D_\nu \Phi^i   
+ \frac 12 e^{-\sigma}\, 
[\Phi^i ,\Phi^j ][\Phi^{i'} ,\Phi^{j'} ] \d_{ii'} \d_{jj'}) , \nn\\
\label{S-phi-effective-nonabelian}
\eea
where $D_\mu \Phi^i = \partial_\mu \Phi^i + i [A_\mu,\Phi^i]$. 
We will indeed verify that the corresponding
equations of motion follow from \eq{matrix-eom}.
It is worth pointing out that in the case $D=10$ and upon
adding suitable fermions as in the IKKT model \cite{Ishibashi:1996xs}, 
this will lead to the analog of $N=4$ SYM theory, albeit on a general
background geometry. The SUSY transformations then 
act non-trivially on the geometry, cf. the discussion in 
\cite{Klammer:2008df}.

\paragraph{Equations of motion.}

We now derive the equations of motion for the gauge field $A_\mu$
from the effective action \eq{action-expanded-2}. Consider first 
\bea
\d S_{NC,2} &=&
- \int \rho \eta ((\d F\theta) (F\theta) - 2 tr(\d F \theta F \theta) \nn\\
&=& 2\int \d A_\nu D_\mu \Big(
\rho \eta F_{\sigma\eta}
(\theta^{\mu\nu} \theta^{\sigma\eta}
 - 2  \theta^{\nu\sigma} \theta^{\eta\mu} )\Big)\nn\\
&=&  2 \int  \rho \d A_\nu F_{\sigma\eta}\, \partial_\mu \eta
( \theta^{\mu\nu} \theta^{\sigma\eta}
 - 2 \theta^{\eta\mu}\theta^{\nu\sigma})
\eea
where $D_\mu = \partial_\mu +i [A_\mu,.]$,
using the identity $\partial_\mu(\rho \theta^{\mu\nu}) =0$ 
\eq{partial-theta-id},
 the Bianci identities for $F$:
\be
(\theta^{\mu\nu}\theta^{\sigma\eta} - 2 \theta^{\nu\sigma}\theta^{\eta\mu})
D_\mu F_{\sigma\eta} \nn\\
= \theta^{\mu\nu}\theta^{\sigma\eta}
\Big(-D_\sigma F_{\eta\mu} - D_\eta F_{\mu\sigma}\Big)
 - 2 \theta^{\nu\sigma}\theta^{\eta\mu} D_\mu F_{\sigma\eta} =0 ,
\ee 
and the Jacobi identity for $\theta^{\mu\nu}$ which implies
\bea
&& \int \rho \eta \d A_\nu F_{\sigma\eta} 
\Big(\theta^{\mu\nu} \partial_\mu \theta^{\sigma\eta}
 - 2 \theta^{\eta\mu}\partial_\mu\theta^{\nu\sigma} \Big) \nn\\
&=&\int \rho \eta \d A_\nu F_{\sigma\eta} 
\Big(-\theta^{\mu\sigma} \partial_\mu \theta^{\eta\nu}
-\theta^{\mu\eta} \partial_\mu \theta^{\nu\sigma}
 + 2 \theta^{\mu\eta}\partial_\mu\theta^{\nu\sigma} \Big) =0 .
\eea
Now consider the terms $\theta \wedge \hat\theta$
\bea
\d S_{NC,1} &=&
 \int \rho (\d F\theta) (F\hat\theta) - 2 \tr(\d F \theta F \hat\theta)
 + (\d F\hat\theta) (F\theta) - 2 tr(\d F \hat\theta F \theta) \nn\\
&=& 2 \int \d A_\nu D_\mu
\Big(\rho F_{\sigma\eta}(\theta^{\nu\mu} \hat\theta^{\sigma\eta}
 + \hat\theta^{\nu\mu} \theta^{\sigma\eta}
 + 2 \theta^{\nu\sigma} \hat\theta^{\eta\mu}
+ 2\hat\theta^{\nu\sigma}\theta^{\eta\mu})\Big) \nn\\
&=& 2 \int \d A_\nu \tilde\nabla^A_\mu
\Big(\rho F_{\sigma\eta}(\theta^{\nu\mu} \hat\theta^{\sigma\eta}
 + \hat\theta^{\nu\mu} \theta^{\sigma\eta}
 + 2 \theta^{\nu\sigma} \hat\theta^{\eta\mu}
+ 2\hat\theta^{\nu\sigma}\theta^{\eta\mu})\Big) \nn
\eea
where $\tilde\nabla^A_\mu = \tilde\nabla_\mu +i [A_\mu,.]$,
using the fact that
expression in brackets is completely anti-symmetric 
in $\mu\sigma\eta$. Putting this together gives the
e.o.m.
\bea
0 &=& 2\sqrt{|\tilde G|}\,
 \tilde\nabla^A_\rho\, (e^{\sigma} F^{\rho\nu}) 
  + \rho F_{\sigma\eta}\, \partial_\mu \eta
( \theta^{\mu\nu} \theta^{\sigma\eta}
 - 2 \theta^{\eta\mu}\theta^{\nu\sigma}) \nn\\
&& + \tilde\nabla^A_\mu
\Big(\rho F_{\sigma\eta}(\theta^{\nu\mu} \hat\theta^{\sigma\eta}
 + \hat\theta^{\nu\mu} \theta^{\sigma\eta}
 + 2 \theta^{\nu\sigma} \hat\theta^{\eta\mu}
+ 2\hat\theta^{\nu\sigma}\theta^{\eta\mu})\Big) .
\label{eom-nonabel}
\eea
This matches precisely with \eq{cons-nonabelian}, which 
is derived from the matrix equations of motion \eq{matrix-eom} 
resp. the conservation law \eq{cons-tang}.

\subsection{Covariant Yang-Mills equations from matrix model}

Consider the matrix equations of motion
for
\be
Y^\mu = X^\mu + \cA^\mu = 
X^\mu - \theta^{\mu\nu}(x) A_\nu
\ee
where $A_\nu = A_{\nu,\a} \lambda^\a$ is a $su(n)$ gauge field,
setting $\Phi^i =0$ but keeping  $\phi^i$.
Then
\bea
-i [Y^\mu,Y^\nu] &=& \theta^{\mu\nu} 
+ \theta^{\mu\rho}\partial_\rho \cA^{\nu}
- \theta^{\nu\rho}\partial_\rho \cA^{\mu} -i [\cA^{\mu},\cA^{\nu}] \nn\\
&=&  (1+\cA\cdot \partial)\theta^{\mu\nu} 
+  \cF^{\mu\nu} 
\label{XX-gauge}
\eea 
where
\be
\cF^{\mu\nu} = - \theta^{\mu\mu'} \theta^{\nu\nu'}F_{\mu'\nu'}
\ee
denoting $\cA\cdot \partial\phi \equiv \cA^\rho\partial_\rho \phi$.
Furthermore, we note the following useful formulae
\bea 
[Y,\phi^i +  \cA^\rho\partial_\rho \phi^i]
&=& [Y,Y^\rho] \partial_\rho \phi^i+ i\cA^\rho \{Y,\partial_\rho \phi^i\}
\label{phi-A-commutator-1}\\[1ex]
\, [Y^\mu,\phi^i +  \cA^\rho\partial_\rho \phi^i]
&=& [Y^\mu,Y^\rho] \partial_\rho \phi^i+ i\cA^\rho \{Y^\mu,\partial_\rho \phi^i\} \nn\\
&=& i((1+\cA\cdot \partial)\theta^{\mu\rho})\partial_\rho \phi^i
- i\theta^{\mu\mu'} \theta^{\rho\rho'}F_{\mu'\rho'} 
\partial_\rho \phi^i
+ i\cA^\rho \theta^{\mu\eta}\partial_\eta\partial_\rho \phi^i  \nn\\
&=& i(1+\cA\cdot \partial)(\theta^{\mu\rho}\partial_\rho \phi^i)
+ i\cF^{\mu\rho} \partial_\rho \phi^i
\label{phi-A-commutator} \\[1ex]
\,[(1+ \cA^\eta\partial_\eta) \phi^i,(1+ \cA^\rho\partial_\rho) \phi^j]
&=& [(1+ \cA^\eta\partial_\eta) \phi^i,Y^\rho] \partial_\rho \phi^j
+ i\cA^\rho \{(1+ \cA^\eta\partial_\eta) \phi^i,\partial_\rho \phi^j\}
\nn\\
&=& -\(i(1+\cA\cdot \partial)(\theta^{\rho\eta}\partial_\eta \phi^i)
+ i\cF^{\rho\eta} \partial_\eta \phi^i\) \partial_\rho \phi^j
+ i\cA^\rho \{\phi^i,\partial_\rho \phi^j\}\nn\\
&=& i(1+\cA\cdot \partial)
(\theta^{\eta\rho} \partial_\eta \phi^i \partial_\rho \phi^j)
 +i\cF^{\eta\rho} \partial_\eta \phi^i \partial_\rho \phi^j
\eea
which hold to $O(\theta^2)$.
We can now obtain the commutative limit of $T^{ab}$ 
\eq{T-ab} for nonabelian gauge fields:
\bea
T^{\mu\nu} &=& [Y^\mu,Y^\rho][Y^\nu,Y^{\rho'}]\eta_{\rho\rho'} 
  + [Y^\nu,Y^\rho][Y^\mu,Y^{\rho'}] ]\eta_{\rho\rho'}
  - \frac 12 g^{\mu\nu} [Y^\rho,Y^\sigma][Y^{\rho'},Y^{\sigma'}]
\eta_{\rho\rho'} \eta_{\sigma\sigma'} \nn\\
 && + [Y^\mu,(1+\cA\cdot\partial)\phi_r][Y^\nu,(1+\cA\cdot\partial)\phi_r]
 + [Y^\nu,(1+\cA\cdot\partial)\phi_r][Y^\mu,(1+\cA\cdot\partial)\phi_r] \nn\\
&&  - \eta^{\mu\nu} [Y^\rho,(1+\cA\cdot\partial)\phi_r]
  [Y^{\rho'},(1+\cA\cdot\partial)\phi_r] \eta_{\rho\rho'} \nn\\
&& - \frac 12 \eta^{\mu\nu} [(1+\cA\cdot\partial)\phi_r,(1+\cA\cdot\partial)\phi_s]
  [(1+\cA\cdot\partial)\phi_r,(1+\cA\cdot\partial)\phi_s]  \nn\\
&\sim& (1+\cA\cdot \partial) 
(- 2  G^{\mu\nu} + 2 \eta^{\mu\nu} \eta(y)) \nn\\
 && - 2 \cF^{\mu\rho} \theta^{\nu\rho'}\eta_{\rho\rho'} 
 -2 \cF^{\mu\rho}\theta^{\nu\rho'} \d g_{\rho\rho'}  
 - 2 \theta^{\mu\rho} \cF^{\nu\rho'}\eta_{\rho\rho'} 
  -2 \theta^{\mu\rho}\cF^{\nu\rho'} \d g_{\rho\rho'}   \nn\\
 &&  +  \eta^{\mu\nu} \( \eta_{\rho'\sigma'}\theta^{\rho\rho'} \cF^{\sigma\sigma'}
   \eta_{\rho\sigma}
+ 2  \theta^{\rho\rho'} \cF^{\sigma\sigma'} \eta_{\rho'\sigma'}
  \d g_{\rho\sigma}
 +  \d g_{\rho'\sigma'}\theta^{\rho\rho'} \cF^{\sigma\sigma'} \d g_{\rho\sigma}\) \nn\\
&=& (1+\cA\cdot \partial) 
(- 2  G^{\mu\nu} + 2 \eta^{\mu\nu} \eta(y)) \nn\\
 && - 2 \cF^{\mu\rho} \theta^{\nu\rho'}g_{\rho\rho'}(x) 
 - 2 \theta^{\mu\rho} \cF^{\nu\rho'}g_{\rho\rho'}(x) 
 +   \eta^{\mu\nu} ( g_{\rho'\sigma'}(x)\theta^{\rho\rho'} \cF^{\sigma\sigma'}
   g_{\rho\sigma}(x)) \nn\\
&=& (1+\cA\cdot \partial) 
(- 2  G^{\mu\nu} + 2 \eta^{\mu\nu} \eta(y)) \,
 - 2 \cG^{\mu\nu} + \frac 12 \eta^{\mu\nu} (\cG^{\a\b} g_{\a\b}) \nn\\[1ex]
T^{\mu r} &=& 
- [Y^\mu,Y^\rho][Y^{\rho'},(1+\cA\cdot\partial)\phi_r]\eta_{\rho\rho'} 
  - [Y^\rho,(1+\cA\cdot\partial)\phi_r][Y^\mu,Y^{\rho'}] \eta_{\rho\rho'} \nn\\
&& + [Y^\mu,(1+\cA\cdot\partial)\phi_t][(1+\cA\cdot\partial)\phi_r,(1+\cA\cdot\partial)\phi_{t'}]\d_{tt'} \nn\\
&& + [(1+\cA\cdot\partial)\phi_r,(1+\cA\cdot\partial)\phi_{t'}] [Y^\mu,(1+\cA\cdot\partial)\phi_t] \d_{tt'} \nn\\
 &\sim& - 2(1+\cA\cdot \partial) \theta^{\rho\rho'} \theta^{\mu\mu'} g_{\rho'\mu'}(y)
 \partial_{\rho}\phi_r 
- 2(\cF^{\rho\rho'} \theta^{\mu\mu'}  g_{\rho'\mu'}(x)
+\theta^{\rho\rho'} \cF^{\mu\mu'} g_{\rho'\mu'}(x))\partial_{\rho}\phi_r \nn\\
&=&  -2(1+\cA\cdot \partial) (G^{\rho\mu}\partial_{\rho}\phi_r)
- 2 \cG^{\mu\nu}\partial_{\rho}\phi_r   
\eea
where we introduce the ``nonabelian metric''
\bea
\cG^{\mu\nu} &=& \theta^{\mu\a}F_{\a\b} G^{\b\nu}
 + \theta^{\nu\a}F_{\a\b}G^{\mu\b} = \cG^{\nu\mu}
\label{nonabel-metric}
\eea
and note that
\be
\hat\theta^{\mu\nu} F_{\mu\nu} 
= - \frac 12 \cG^{\mu\nu}g_{\mu\nu} .
\ee
This gives
\bea
 -i[Y^{\mu'},T^{\mu \nu}]\eta_{\mu'\mu} 
&=& -i[Y^{\mu'},(1+\cA\cdot \partial)(- 2  G^{\mu\nu} + 2\eta^{\mu\nu} \eta(y))]\eta_{\mu'\mu}  \nn\\
&& - \theta^{\mu'\rho}(\partial_\rho + i [A_\rho,.])
\Big(2\cG^{\mu\nu} - \frac 12 \eta^{\mu\nu} (\cG^{\a\b} g_{\a\b}) \Big)
\eta_{\mu'\mu}\nn\\
&=& (1+\cA\cdot \partial)\(\theta^{\mu'\rho}\partial_\rho (- 2  G^{\mu\nu} + 2\eta^{\mu\nu} \eta(y))\)\eta_{\mu'\mu} 
+ \cF^{\mu'\rho} \partial_\rho (- 2  G^{\mu\nu} + 2\eta^{\mu\nu} \eta(y))\eta_{\mu'\mu} \nn\\
&& - \theta^{\mu'\rho}(\partial_\rho + i [A_\rho,.])
\Big( 2\cG^{\mu\nu} - \frac 12 \eta^{\mu\nu} (\cG^{\a\b} g_{\a\b})\Big)\eta_{\mu'\mu}
\eea 
and
\bea
i\frac 12 [(1+\cA\cdot\partial)\phi_r,T^{r \nu}] 
&=& (1+\cA\cdot \partial)\(\theta^{\eta\rho} 
\partial_\eta \phi_r \partial_\rho (G^{\mu\nu}\partial_{\mu}\phi_r)\)
 +\cF^{\eta\rho} \partial_\eta \phi_r \partial_\rho (G^{\mu\nu}\partial_{\mu}\phi_r)
\nn\\ 
&& - i [(1+\cA\cdot\partial)\phi_r,\cG^{\nu\mu}\partial_{\mu}\phi_r  ] \nn\\
&=& (1+\cA\cdot \partial)\(\theta^{\eta\rho} 
\partial_\rho ( G^{\mu\nu}\d g_{\eta\mu})\)
 +\cF^{\eta\rho} \partial_\rho (G^{\mu\nu}\d g_{\eta\mu})
 -i [Y^\sigma,\cG^{\nu\rho}\partial_{\rho}\phi_r]\partial_\sigma \phi_r\nn\\ 
&=& (1+\cA\cdot \partial)\(\theta^{\eta\rho} 
\partial_\rho ( G^{\mu\nu}\d g_{\eta\mu})\)
 +\cF^{\eta\rho} \partial_\rho (G^{\mu\nu}\d g_{\eta\mu})
 + \theta^{\sigma\eta} D_\eta
\(\cG^{\nu\mu}\d g_{\mu\sigma}\) \nn
\eea
to $O(\theta^2)$, 
where $D_\mu = \partial_\mu + i [A_\mu,.]$, using
\eq{phi-A-commutator-1}.
Hence the ``tangential'' conservation law \eq{cons-tang} gives
\bea
0 &=& -i[Y^{\mu'},T^{\mu \nu}]\eta_{\mu'\mu} 
 -i [Y^r,T^{r \nu}] \nn\\
&=& 2(1+\cA\cdot \partial)\(\theta^{\nu\rho}\partial_\rho \eta(y)
-\theta^{\eta\rho} \partial_\rho ( G^{\rho\nu} g_{\eta\rho}) \) 
-2 \cF^{\eta\rho} \partial_\rho ( G^{\mu\nu} g_{\eta\mu})
+ 2 \cF^{\nu\rho} \partial_\rho\eta(y) \nn\\
&& - \theta^{\eta\rho} D_\rho
\Big( 2\cG^{\mu\nu} g_{\eta\mu} - \frac 12 \d^\nu_\eta(\cG^{\a\b} g_{\a\b})\Big) .
\label{cons-tang-explicit-nonabelian}
\eea 
Using the geometrical conservation law \eq{cons-tang-explicit-1}, 
this reduces to
\bea
0 &=& -\cF^{\eta\rho} \partial_\rho (G^{\mu\nu} g_{\eta\mu})
+\cF^{\nu\rho} \partial_\rho\eta(y) 
 - \theta^{\eta\rho} D_\rho 
\Big( \cG^{\mu\nu} g_{\eta\mu} 
  - \frac 14 \d^\nu_\eta(\cG^{\a\b} g_{\a\b})\Big)  
\label{cons-tang-explicit-nonabelian-1}
\eea
Now we note that 
\bea
\theta^{\eta\rho} D_\rho (\cG^{\mu\nu} g_{\eta\mu}) 
&=&-\rho^{-1} D_\rho
(\rho\theta^{\rho\eta}g_{\eta\mu}\cG^{\mu\nu} ) \nn\\
&=& -\rho^{-1} D_\rho (-\rho G^{\rho\a}F_{\a\b} G^{\b\nu}
 + \rho \hat\theta^{\rho\b}F_{\b\a}\theta^{\a\nu}) \nn\\
&=& \rho^{-1}\sqrt{|\tilde G|}\,(\tilde\nabla_\rho + i [A_\rho,.])
(e^{\sigma} F^{\rho\nu})
 -\rho^{-1} D_\rho(\rho \hat\theta^{\rho\b}F_{\b\a}\theta^{\a\nu}). \nn\\
\eea
Similarly, 
\be
\frac 14\theta^{\nu\rho} D_\rho (\cG^{\a\b} g_{\a\b})
= -\frac 12\theta^{\nu\rho} D_\rho (\hat \theta^{\a\b} F_{\a\b})
= -\frac 12\rho^{-1} D_\rho 
 (\rho\theta^{\nu\rho} \hat \theta^{\a\b} F_{\a\b})
\ee
and
\bea
\rho\cF^{\eta\rho} \partial_\rho (G^{\mu\nu} g_{\eta\mu})
&=& 
\theta^{\eta\a} F_{\a\b} 
\partial_\rho (\rho\theta^{\b\rho} G^{\mu\nu} g_{\eta\mu}) \nn\\
&=&  F_{\a\b} 
\partial_\rho (\rho\theta^{\b\rho}G^{\nu\mu} g_{\mu\eta}\theta^{\eta\a})
- \rho F_{\a\b} g_{\eta\mu} G^{\mu\nu} 
 \theta^{\b\rho}\partial_\rho \theta^{\eta\a}\nn\\
&=& F_{\a\b} 
\partial_\rho (\rho\theta^{\b\rho} \hat \theta^{\nu\a})
+\frac 12 \rho F_{\a\b} g_{\eta\mu} G^{\mu\nu} 
 \theta^{\eta\rho}\partial_\rho \theta^{\a\b}\nn\\
&=& F_{\a\b} 
\partial_\rho (\rho\theta^{\b\rho} \hat \theta^{\nu\a})
+\frac 12 \rho F_{\a\b} \hat\theta^{\nu\rho}
\partial_\rho \theta^{\a\b}\nn\\
&=&  F_{\a\b} 
\partial_\rho \Big(\rho\theta^{\b\rho} \hat \theta^{\nu\a}
+\frac 12 \rho\hat\theta^{\nu\rho}\theta^{\a\b}\Big)
-\frac 12 \rho F_{\a\b}\theta^{\nu\rho}
 \theta^{\a\b} \partial_\rho\eta \nn\\
&=&  D_\rho \Big(\rho F_{\a\b} 
(\theta^{\b\rho} \hat \theta^{\nu\a}
+\frac 12 \hat\theta^{\nu\rho}\theta^{\a\b})\Big)
-\frac 12 \rho F_{\a\b}\theta^{\nu\rho}
 \theta^{\a\b} \partial_\rho\eta
\eea
since
$$
2 F_{\a\b} \theta^{\b\rho}\partial_\rho \theta^{\eta\a}
= -F_{\a\b} \theta^{\eta\rho}\partial_\rho \theta^{\a\b}
$$
using the Jacobi identity, and the Bianci identity
$$
 D_\rho F_{\a\b} 
\Big(\theta^{\b\rho} \hat \theta^{\nu\a}
+\frac 12 \hat\theta^{\nu\rho}\theta^{\a\b}\Big) 
= -\hat \theta^{\nu\a}\theta^{\b\rho}  
(D_\a F_{\b\rho}+ D_\b F_{\rho\a}+ D_\rho F_{\a\b})
=0\, .
$$
Therefore \eq{cons-tang-explicit-nonabelian-1} becomes
\bea
0 &=& -\rho^{-1}\sqrt{|\tilde G|}\,(\tilde\nabla_\rho + i [A_\rho,.])
(e^{\sigma} F^{\rho\nu}) 
 + \frac 12 F_{\a\b} (2\theta^{\nu\a}\theta^{\b\rho}
+ \theta^{\nu\rho} \theta^{\a\b}) \partial_\rho\eta \nn\\
&& +\frac 12\rho^{-1} D_\rho
\Big(\rho F_{\a\b} (2\theta^{\nu\a}\hat\theta^{\rho\b} 
 +\theta^{\nu\rho} \hat \theta^{\b\a} 
 + 2\hat\theta^{\nu\a}\theta^{\rho\b} 
+ \hat\theta^{\nu\rho}\theta^{\b\a})\Big) \nn\\
 &=& -\rho^{-1}\sqrt{|\tilde G|}\,
 \tilde\nabla_\rho^A (e^{\sigma} F^{\rho\nu}) 
 + \frac 12 F_{\a\b} (2\theta^{\nu\a}\theta^{\b\rho}
 +\theta^{\nu\rho} \theta^{\a\b}) \partial_\rho\eta\nn\\
 && +\frac 12\rho^{-1} \tilde\nabla_\rho^A 
 \Big(\rho F_{\a\b} (2\theta^{\nu\a}\hat\theta^{\rho\b} 
  +\theta^{\nu\rho} \hat \theta^{\b\a} 
  + 2\hat\theta^{\nu\a}\theta^{\rho\b} 
 + \hat\theta^{\nu\rho}\theta^{\b\a})\Big) 
\label{cons-nonabelian}
\eea
where $\tilde\nabla_\mu^A = \tilde\nabla_\mu + i [A_\mu,.]$,
using the fact that
expression in brackets is completely anti-symmetric in $\a\b\rho$.
This is precisely the e.o.m. \eq{eom-nonabel}
derived from the effective action \eq{action-expanded-2}.

It is remarkable that this e.o.m. 
is derived from an underlying
symmetry of the model. This means that it is protected from quantum
corrections. In particular, the coefficients of the would-be
topological term is fixed and cannot change under renormalization. 
While this is formally true even in the $D=4$ model, the quantization 
probably makes sense only in the $D=10$ model.

One might object that the equations of motion do not determine 
or rule out a ``truly-topological'' term $\sim \int  F \wedge F$. 
However, such a term could come in the $D=10$ model only through
$\int \rho FF \theta\theta$, which has the wrong scaling dimension
in $\theta$
and is therefore excluded. Hence the action is determined uniquely 
to be of the form \eq{action-expanded-2}, \eq{S-NC-def}.

This mechanism for determining the would-be topological action is
quite remarkable and should have profound consequences 
once a semi-realistic vacuum with non-trivial 
low-energy gauge group is found.
This can also be seen in connection with the finiteness
of the $N=4$ SYM theory, where the so-called $\theta$-angle
is protected from running under renormalization
by supersymmetry; this in turn is related to 
Montonen-Olive duality \cite{Montonen:1977sn,Vafa:1994tf}.

\paragraph{Nonabelian scalar fields.}

Finally, 
the action \eq{S-phi-effective-nonabelian} implies
the following equations of motion for the nonabelian scalars
\be
0 =  \tilde G^{\mu\nu}
(\tilde\nabla_\nu + i [A_\nu,.]) D_\mu \Phi^k  
+ e^{-\sigma}\, [\Phi^i ,[\Phi^k ,\Phi^{j} ]] \d_{ij} . 
\label{eom-nonabelian}
\ee
This e.o.m. is easily derived from the basic 
 matrix e.o.m. \eq{matrix-eom}, which
is simply a nonabelian version of \eq{eom-varphi-0}:
keeping only contributions up to $O(\theta^2)$, we have
\bea
[Y^a, [Y^{b},\Phi^k]] \eta_{ab} &=& 
[Y^\mu, [Y^{\nu},\Phi^k]]\eta_{\mu\nu} 
 + [\phi^i\otimes \one_n + \Phi^i, 
[\phi^j\otimes \one_n +\Phi^j,\Phi^k]]\,\d_{ij}\nn\\
&\sim& - \theta^{\mu\rho}
\(\partial_\rho + i [A_{\rho},.]\)
\(\theta^{\nu\eta}(\partial_\eta + i [A_\eta,.])
\Phi^k \) \eta_{\mu\nu} \nn\\
&& - \{\phi^i,\theta^{\nu\eta}\partial_\nu\phi^j\partial_\eta\Phi^k\}\,\d_{ij} 
 +  [\Phi^i, [\Phi^j,\Phi^k]]\,\d_{ij} \nn\\
&=&  - \theta^{\mu\rho}
\(\partial_\rho + i [A_{\rho},.]\)
\(\theta^{\nu\eta}(\partial_\eta + i [A_\eta,.])\Phi^k\) \eta_{\mu\nu} \nn\\
&& - \rho^{-1}\partial_\rho\Big(\rho\theta^{\mu\rho}
\theta^{\nu\eta}\d g_{\mu\nu}\partial_\eta\Phi^k\Big)
 +  [\Phi^i, [\Phi^j,\Phi^k]]\,\d_{ij}\nn\\
&=&  - G^{\rho\eta}
(\partial_\rho + i [A_{\rho},.])
(\partial_\eta + i [A_\eta,.])\Phi^k \nn\\
&& - e^\sigma \tilde\Gamma^\eta
(\partial_\eta + i [A_\eta,.])\Phi^k 
+  [\Phi^i, [\Phi^j,\Phi^k]]\,\d_{ij}  \nn\\
&=&  - e^\sigma  \tilde G^{\rho\eta}
(\tilde\nabla_\rho + i [A_{\rho},.])
(\partial_\eta + i [A_\eta,.])\Phi^k 
+  [\Phi^i, [\Phi^j,\Phi^k]]\,\d_{ij}  \nn
\eea
using \eq{eom-extradim}.
This involves the expected gauge-covariant Laplacian. 
It would be interesting to compute also higher-order corrections
in $\theta$, which might involve curvature contributions.

In principle, the e.o.m. for the nonabelian gauge fields
\eq{cons-nonabelian} could also be derived in a similar way. However
this is manageable only in normal embedding coordinates
\eq{normal-embedd-coords}, i.e. taking advantage of the underlying 
$SO(D)$ symmetry of the model. The derivation presented 
above based on Noethers theorem is not only more profound but 
also considerably simpler.

\section{Discussion and conclusion}

In this paper, the foundations of the matrix-model
approach to (emergent) gravity are developed further. 
The effective action for nonabelian gauge fields
on non-trivially embedded branes 
$\cM_\theta\subset \R^D$ is obtained, which turns out to be the expected 
generalization of the result in \cite{Steinacker:2007dq} 
for flat embeddings. This shows that 
the effective metric $\tilde G_{\mu\nu}$ \eq{G-tilde-def} governs also
the nonabelian gauge fields, and must therefore be interpreted
as gravitational metric. Since $\tilde G_{\mu\nu}$ 
depends on the embedding metric $g_{\mu\nu}$, 
we can indeed obtain the most general effective
metric in this framework for $D \geq 10$
using standard embedding theorems \cite{friedman}, at least locally.
Therefore the $D=10$ matrix model, notably the IKKT model,
is in principle capable of describing generic 4-dimensional metrics.
The mechanism of emergent gravity under consideration here 
becomes a serious candidate for a realistic theory of gravity.

Furthermore, we have shown that the equations of motion for 
the Poisson tensor $\theta^{\mu\nu}$ and the 
nonabelian gauge fields are consequences of a Noether theorem,
and therefore protected from quantum corrections. This is 
very remarkable, and supports the viability of this framework
at the quantum level. 
We also elaborated in some more detail the metric 
fluctuations around the flat Moyal-Weyl background.
This leads to a simplified understanding of the Ricci-flatness
of the $U(1)$ modes as first observed in \cite{Rivelles:2002ez}, 
interpreted as gravitational waves. 
On the other hand, one of the main open problems 
is to find an analog of the Schwarzschild solution.
Since the Einstein-Hilbert action is induced only upon quantization,
this can be addressed reliably only once the one-loop effective
action is known. However, the validity of the geometric
equation \eq{eom-geom-covar-extra} at the quantum level established here
should be a useful guide already at this point.

Even though the physical properties of emergent gravity are not 
well understood, there are several striking aspects 
which make this framework very attractive.
First, its definition requires no classical-geometrical 
notions of geometry whatsoever. The geometry arises
dynamically, and in fact physical condensation and 
melting processes of the geometry
have been observed in similar toy models 
\cite{DelgadilloBlando:2008vi}. This is very appealing from
the point of view of quantum gravity and cosmology. 
Another fascinating aspect of the present framework
is that it leads to a unified picture 
of gravitons and nonabelian gauge fields, which arise as
abelian resp. nonabelian fluctuations of the basic matrices
(covariant coordinates) around a geometrical background.
The quantization around a such a background
is technically rather straightforward, similar
to nonabelian gauge field theory. Remarkably, flat space 
remains to be a solution at one loop, 
cf. \cite{Steinacker:2007dq,Steinacker:2008ri} and a related
discussion in \cite{Yang:2008fb}.
In this context, further analytical and numerical work
on the emergence of 4-dimensional NC branes in matrix
models is needed, particularly for the supersymmetric case; see also
\cite{Nishimura:2001sx,Azeyanagi:2008bk} for 
work in this context.

We conclude by stressing
that the  gravity ``theory'' under consideration is not based on 
some theoretical expectations, but is simply the result of a careful 
analysis of the semi-classical limit of this type of matrix models.
Clearly much more work is required,
and it seems likely that many more surprising 
results and insights are waiting to be 
discovered.

\paragraph{Acknowledgments}

I would like to thank in particular H. Grosse
for useful discussions and support, and P. Aschieri, L. Freidel,
D. Klammer, K. Krasnov, J. Madore, 
M. Buric,  P. Schreivogl and P. Schupp 
for discussions on various related topics. 
This work was supported by the FWF project P20017.

\end{document}